\documentclass[aps,prd,preprintnumbers,groupedaddress,nofootinbib]{revtex4}

\pdfoutput=1
\usepackage{subfigure}
\usepackage{graphicx}
\usepackage{aas_macros}
\usepackage{amsfonts}
\usepackage{amssymb}
\usepackage{amsmath}
\usepackage{slashed}
\usepackage{hyperref}
\usepackage{url}
\usepackage{multirow}
\usepackage{color}
\definecolor{darkblue}{RGB}{0,0,175}

\newcommand{\be}{\begin{equation}}
\newcommand{\ee}{\end{equation}}
\newcommand{\ba}{\begin{eqnarray}}
\newcommand{\ea}{\end{eqnarray}}
\begin{document}
\include{declare}

\title{Scattering, Damping, and Acoustic Oscillations: \\ Simulating the Structure of Dark Matter Halos with Relativistic Force Carriers}

\author{Matthew R.~Buckley$^{1}$, Jes\'{u}s Zavala$^{2}$\footnote{Marie Curie Fellow}, Francis-Yan Cyr-Racine$^{3,4}$\footnote{Keck Institute for Space Studies Fellow}, Kris Sigurdson$^{5}$, and Mark Vogelsberger$^6$}
\affiliation{$^1$Department of Physics and Astronomy, Rutgers University, Piscataway, NJ 08854, USA}
\affiliation{$^2$Dark Cosmology Centre, Niels Bohr Institute, University of Copenhagen, Juliane Maries Vej 30, 2100 Copenhagen, Denmark}
\affiliation{$^3$NASA Jet Propulsion Laboratory, California Institute of Technology, Pasadena, CA 91109, USA}
\affiliation{$^4$California Institute of Technology, Pasadena, CA 91125, USA}
\affiliation{$^5$Department of Physics and Astronomy, University of British Columbia, Vancouver, BC, V6T 1Z1, Canada}
\affiliation{$^6$Department of Physics, Kavli Institute for Astrophysics and Space Research, Massachusetts Institute of Technology, Cambridge, MA 02139, USA}

\preprint{}
\date{\today}

\begin{abstract}
We demonstrate that self-interacting dark matter models with interactions mediated by light particles can have significant deviations in the matter power-spectrum and detailed structure of galactic halos when compared to a standard cold dark matter scenario. While these deviations can take the form of suppression of small scale structure that are in some ways similar to that of warm dark matter, the self-interacting models have a much wider range of possible phenomenology. A long-range force in the dark matter can introduce multiple scales to the initial power spectrum, in the form of dark acoustic oscillations and an exponential cut-off in the power spectrum. Using simulations we show that the impact of these scales can remain observationally relevant  up to the present day. Furthermore, the self-interaction can continue to modify the small-scale structure of the dark matter halos, reducing their central densities and creating a dark matter core. The resulting phenomenology is unique to this type of models.
\end{abstract}
\maketitle

\section{Introduction \label{sec:intro}}

Dark matter is one of the clearest signals of the existence of new physics beyond the Standard Model of particle physics. The agreement between observations and the predictions of cold dark matter with a cosmological constant ($\Lambda$CDM) \cite{Ade:2013zuv} gives solid evidence of the gravitational interaction between dark matter and the Standard Model baryons. Astrophysical observations provide only one further piece of information about the nature of dark matter: namely that it is ``cold,'' or at most ``warm,'' in that it cannot have been highly relativistic during structure formation in the Early Universe (see {\it e.g.}~\cite{Bertone:2004pz}). However, discrepancies between simulations and observations of dark matter structure at small scales \cite{Klypin:1999uc,Moore:1999nt,Spergel:1999mh,Zavala:2009,Oh:2010ea,BoylanKolchin:2011dk,BoylanKolchin:2011de,Walker:2011zu,2013MNRAS.433.3539G,Weinberg:2013aya,Kirby:2014sya,Tollerud:2014zha,Garrison-Kimmel:2014vqa} have led to an interest in models with large dark matter self-scattering cross sections. Though it is possible that baryonic physics can resolve the issue \cite{Pontzen:2011ty,Governato:2012fa,Teyssier:2012ie,Brooks:2012vi,Brooks:2012ah,Zolotov:2012xd,Arraki:2012bu,Amorisco:2014}, such models can provide a rich range of phenomenology beyond those exhibited in cold dark matter (CDM) and so are a subject of much theoretical and observational interest.

Given the large cross sections that would be required to significantly alter the halo properties of dark matter through self-interactions (comparable to the nuclear cross sections of the visible sector), it is necessary to reconsider the early Universe cosmology realized in such models. While the cold dark matter would still decouple from the standard model plasma as in a typical dark matter scenario, the self-interactions keep the dark matter in kinetic equilibrium with itself until comparatively late in the Universe's evolution. If these interactions are mediated by a light (relativistic) force carrier whose energy density is large compared to that of dark matter, then the sound speed in the dark matter fluid remains high and sound waves can propagate long distances, potentially leaving signatures on cosmological and astrophysical scales. This can also occur if the self-interaction is mediated by a force carrier that also couples dark matter to a Standard Model relativistic species, such as the neutrinos \cite{Mangano:2006mp,Boehm:2006mi,Hooper:2007tu,Serra:2009uu,Aarssen:2012fx,Diamanti:2012tg,Tulin:2013teo,Boehm:2013jpa,Shoemaker2013157,Laha:2013xua}. Though not all self-interacting dark matter (SIDM) models possesses such a light particle, it is not uncommon \cite{Goldberg:1986nk,Hodges:1993yb,Berezhiani:1995am,Berezhiani:1995yi,Foot:1995pa,Mohapatra:2000qx,Berezhiani:2003xm,Foot:2004pa,Foot:2007iy,Pospelov:2007mp,Hooper:2008im,Feng:2008ya,Feng:2008mu,Pospelov:2008jd,Kaplan:2009de,Shepherd:2009sa,Behbahani:2010xa,Kaplan:2011yj,Cline:2012is,CyrRacine:2012fz,Cline:2013pca,Cline:2013zca,Petraki:2014uza,Cline:2014eaa,Foot:2013vna,Foot:2014mia,Boehm:2001hm,Dubovsky:2003yn,Melchiorri:2007sq,McDermott:2010pa,Berezhiani:2012ru,Dolgov:2013una,Wilkinson:2013kia,Dvorkin:2013cea,1992ApJ...398...43C,Gradwohl:1992ue,1994ApJ...431...41M,1995ApJ...452..495D,Ackerman:mha,Feng:2009mn}. As two of the authors have previously noted \cite{Cyr-Racine:2013fsa}, the large sound speed of the dark matter sector in these models allows for ``dark'' acoustic oscillations (DAOs, in analogy to the more familiar baryon acoustic oscillations) that can leave detectable imprints in the anisotropies of the cosmic microwave background (CMB) and in the distribution of matter and galaxies throughout the Universe.

In addition, collisional damping between the light mediator and the dark matter can erase primordial structure below some critical scale set by the diffusion length of the force carrier at the temperature of kinetic decoupling. To first order, one might naively expect such a model to have a phenomenology qualitatively similar  to warm dark matter (WDM) \cite{Dalcanton:2000hn,Bode:2000gq,Avila-Reese:2001}, resulting primarily in a suppressed number of dark matter halos below some critical mass. This has been pointed out in the context of model-building \cite{Mangano:2006mp,Hooper:2007tu,Kaplan:2009de,Shoemaker:2013tda}, but the full range of phenomenology has not been pursued and very few cosmological simulations have been performed including these effects \cite{Kamada:2013sh,Boehm:2014vja}. As we demonstrate in this paper, models with long range forces have a number of unique properties beyond the exponential suppression of small-scale structure. These differences can lead to significant deviations from both collisionless CDM and WDM predictions. Indeed, a SIDM model with light mediators can have novel effects on the small-scale matter power spectrum, the dark matter halo mass function, and the internal structure of surviving halos. We will demonstrate the range of possible behavior of SIDM models with a light force carrier using cosmological $N$-body simulations where the initial conditions are allowed to deviate from that of CDM, and include the effects of momentum exchange through self-interactions on the internal structure of the dark matter halos.

In the next section, we discuss the physics of SIDM with a relativistic force carrier in the formation and evolution of galactic structure from the early Universe till the present. We will focus on three effects:
\begin{itemize}
\item the exponential suppression of primordial structure from collisional damping,

\item the imprint of dark acoustic oscillations in the initial matter power spectrum, and

\item the modification of dark matter halo properties due to the self-interaction of dark matter.
\end{itemize}
This section will also introduce the details of the ``dark atom'' model we adopt as a specific example of a SIDM model with a long-range mediator. In Section~\ref{sec:simulation}, we outline the cosmological simulation methods used as well as the parameters of the dark atom, cold dark matter, and warm dark matter benchmark models. The results of convergence tests for our simulations are included in the Appendix.  In Section~\ref{sec:dao} we show the results of our simulations, demonstrating the present-day effects of both the exponential cut-off and the DAOs in the matter power spectrum. This provides a sampling of the range of possible effects from DAO, as the dark matter microphysics is varied. Finally in Section~\ref{sec:density}, we demonstrate the modifications of the dark matter halo structure due to the both the self-interactions and the changes in the initial power spectrum. Though this paper does not cover the full set of available phenomenology that can arise from SIDM models with long range forces, the range of simulations provides a representative sample of what is possible.

\section{Self-Interactions and Structure Formation \label{sec:preliminaries}}

Though the theory space of dark matter candidates is vast, most of the proposed dark matter particles have extremely small interactions in the present Universe, both internally and between themselves and the baryons. While it is an experimental fact \cite{Akerib:2013tjd} that the dark matter-baryon spin-independent scattering cross sections at Galactic velocities ($\sim 200$~km/s) are constrained to be less than $10^{-43}-10^{-45}$~cm$^2$ for dark matter with masses between $\sim 10-10^4$~GeV, the limits we can place on the self-interactions of dark matter are far weaker. From the separation between the baryonic gas and dark matter in the Bullet Cluster (a system of colliding galaxy clusters), the self-scattering cross section per unit mass must be less than $\sim 1$~cm$^2$/g for relative velocities of order 1000~km/s \cite{Clowe:2006eq,Randall:2007ph}. For a 100~GeV dark matter particle (a representative mass for a thermal neutralino), this is an upper bound of only $\sim 2 \times 10^{-22}$~cm$^2$. This is orders of magnitude above the cross sections with nucleons probed by direct detection, and much larger than the annihilation cross section in the early Universe implied by a cold thermal relic ($\sim 10^{-36}$~cm$^2$).

The ``crisis in small scale structure'' \cite{Klypin:1999uc,Moore:1999nt,Spergel:1999mh,Oh:2010ea,BoylanKolchin:2011dk,BoylanKolchin:2011de,Walker:2011zu} has focused attention on SIDM models with cross sections per mass of approximately 1 cm$^2$/g for dark matter with  velocities of 30-60~km/s, sufficient to alter the structure of dark matter halos on the scale of dwarf galaxies (that is, with masses of $10^{9-10}M_\odot$) \citep{Zavala:2012us}. Relatively speaking, this cross section is enormous, comparable to a nuclear scattering cross section in the Standard Model. While it is not required, such a large cross section makes it reasonable to consider the existence of light force carriers in the dark sector. This could occur, for example, in dark atom models interacting through dark photons \cite{Goldberg:1986nk,Kaplan:2009de,Behbahani:2010xa,Kaplan:2011yj,Cline:2012is,CyrRacine:2012fz,Cline:2013pca,Cline:2013zca,Petraki:2014uza,Cline:2014eaa}, in mirror dark matter models \cite{Hodges:1993yb,Berezhiani:1995am,Berezhiani:1995yi,Foot:1995pa,Mohapatra:2000qx,Berezhiani:2003xm,Foot:2004pa,Foot:2007iy,An:2009vq,Foot:2013vna,Foot:2014mia}, or dark matter coupled to Standard Model photons \cite{Boehm:2001hm,Dubovsky:2003yn,Melchiorri:2007sq,McDermott:2010pa,Berezhiani:2012ru,Dolgov:2013una,Wilkinson:2013kia,Dvorkin:2013cea}, neutrinos \cite{Mangano:2006mp,Boehm:2006mi,Hooper:2007tu,Serra:2009uu,Aarssen:2012fx,Diamanti:2012tg,Tulin:2013teo,Boehm:2013jpa,Shoemaker2013157,Laha:2013xua}, or to some other new light particle \cite{1992ApJ...398...43C,Gradwohl:1992ue,1994ApJ...431...41M,1995ApJ...452..495D,Ackerman:mha,Feng:2009mn}.

A SIDM model where the interactions are mediated by a force carrier that is relativistic during the thermal freeze-out in the early Universe can have significant deviations in the present-day galactic structure compared to both non-interacting CDM as well as SIDM without a relativistic mediator. There are three main effects that we will explore in this paper: an exponential suppression of the initial power spectrum of dark matter halos, DAO in the dark matter thermal bath leaving imprints on the initial matter power spectrum as well as the galaxy correlation function, and modifications of the structure of dark matter halos from energy transport made possible by dark matter self-scattering. The importance of each of these effects varies depending on the parameters of the SIDM model. The first and second effects are determined by the mass and couplings of the relativistic force carrier's interaction with the dark matter in the early Universe, while the third effect is due to the self-scattering cross section as the primordial dark matter halos evolve forward in time, combined with delayed formation for small halos resulting in lower densities. We first introduce our benchmark model for SIDM with a relativistic force carrier, the ``dark atom'' model, and then discuss each of the possible effects in turn, and give an estimate of the range of parameters over which they are relevant. 

\subsection{Dark Atoms \label{sec:darkatoms}}

While there are many SIDM scenarios that mediate the self-scattering interaction through a light force carrier, in this paper we will use the dark atom model as a benchmark \cite{Goldberg:1986nk,Kaplan:2009de,Behbahani:2010xa,Kaplan:2011yj,Cline:2012is,CyrRacine:2012fz,Cline:2013pca,Cline:2013zca,Petraki:2014uza,Cline:2014eaa,Hodges:1993yb,Berezhiani:1995am,Berezhiani:1995yi,Foot:1995pa,Mohapatra:2000qx,Berezhiani:2003xm,Foot:2004pa,Foot:2007iy,Foot:2013vna,Foot:2014mia}. In this model, dark matter is composed of two different particles, oppositely charged under a new long-range abelian gauge group $U(1)_D$, mediated by a dark photon $\phi$. The two particles, analogous to the proton and electron, form a bound state at low energies. Depending on the details of the dark atom parameter space, this results in most of the dark matter being neutral under $U(1)_D$, and so avoid possible constraints on $U(1)_D$-charged dark matter from dark magnetic fields in galaxies \cite{Ackerman:mha}. For in depth investigations of dark atom phenomenology, we refer the reader to Refs.~\cite{Cline:2013pca,Cline:2012is,CyrRacine:2012fz}. The full range of behaviors available to these models is much broader than the subject of this paper, so for our benchmark scenarios we only present results relevant to our present work.

The fundamental parameter space of atomic dark matter (ADM) are the masses $m_P$ and $m_E$ of the oppositely charged ``proton'' $P$ and ``electron'' $E$ analogues (by definition, $m_E < m_P$), the coupling parameter $g_D$ (alternatively, the dark fine structure constant $\alpha_D$), and the dark photon mass $m_\phi$ (massless for an unbroken gauge group). We will assume that both of these fundamental particles are fermionic. Though it is possible for ADM models to accommodate both asymmetric and thermal (symmetric) production mechanisms, for simplicity we assume that the dark matter was produced in some asymmetric process \cite{Nussinov:1985xr,Barr:1990ca,Barr:1991qn,Dodelson:1991iv,Kaplan:1991ah,Kuzmin:1996he,Thomas:1995ze,Fujii:2002aj,Hooper:2004dc,Gudnason:2006ug,Kitano:2008tk,Kaplan:2009ag,Cohen:2009fz,Cohen:2010kn,Shelton:2010ta,Davoudiasl:2010am,Buckley:2010ui,Haba:2010bm,Blennow:2010qp,Hall:2010jx,Graesser:2011wi}. This implies that the number densities of $\bar{P}$ and $\bar{E}$ are much lower than that of $P$ and $E$. 

As the temperature of the dark sector decreases, the $P$ and $E$ particles form bound states $H$ analogous to a hydrogen atom. As a result, there are additional derived parameters of the theory: mass of the bound state $m_D$, the binding energy $B_D$, and the temperature of the dark sector $\hat{T}$. For this paper, we will make the assumption that the dark photon mass $m_\phi$ is negligibly small. This allows the dark atoms to retain the familiar atomic structure of standard baryonic hydrogen, albeit with modified energy levels, and facilitate the computation of their thermal and ionization history \cite{CyrRacine:2012fz}.

Recall that the temperature of the dark sector $\hat{T}$ will in general be lower than the temperature $T$ of the visible sector, as the visible sector will be reheated as particles freeze-out of thermal equilibrium. It is $\hat{T}$, rather than $T$, which will set the number density and velocity of the dark matter and dark radiation. If the temperature-dependent number of degrees of freedom in the two sectors are $g_\text{vis}$ and $g_\text{dark}$, and the two sectors were last in chemical equilibrium at the freeze-out temperature $T_{f}$, then:
\begin{equation}
\xi(T) \equiv \frac{\hat{T}}{T} = \left(\frac{g_\text{vis}(T) g_\text{dark}(T_{f})}{g_\text{vis}(T_{f})g_\text{dark}(T)} \right)^{1/3}.
\end{equation}
As $g$ is a strictly increasing function of $T$, $\xi < 1$ unless there are a large number of degrees of freedom available in the dark sector that decouple after the freeze-out of dark matter. The ratio $\xi$ in ADM is therefore set by model-dependent assumptions in the UV-completion of the theory which are beyond the scope of this work.

As we will be interested in the kinetic decoupling of dark radiation from the the massive dark matter in the early Universe, as well as the self-scattering of the dark atoms in the present day, we require several cross sections for this model. As the Universe cools, the dark radiation will remain in kinetic equilibrium with the dark matter by a combination of Compton scattering off of the charged $P$ and $E$, and by Rayleigh scattering off of the $U(1)_D$-neutral dark atoms. The two relevant cross sections are \cite{CyrRacine:2012fz}:
\begin{eqnarray}
\sigma_\text{Compton} & = & \left(1+\frac{m_E^2}{m_P^2}\right)\sigma_\text{Thomson}, \label{eq:sigmacompton} \\
\sigma_\text{Raleigh} & \approx & 32\pi^4 \left(\frac{\hat{T}}{B_D}\right)^4 \sigma_\text{Thomson} \quad(\hat{T}\ll B_D), \label{eq:sigmaraleigh} \\
\sigma_\text{Thomson} & = & \frac{8\pi\alpha_D^2}{3m_E^2}.
\end{eqnarray}
As the Universe cools, the fraction $x_D$ of ionized $E$ and $P$ decreases \cite{CyrRacine:2012fz}, and so the relative importance of the two scattering modes changes. Since Compton scattering scales as $x_D$ it is most important before dark atom recombination, while Raleigh scattering, scaling as $1-x_D$, can only become relevant after the onset of dark recombination. 

To study the effects of particle collisions inside dark matter halos, we also need to determine the elastic self-scattering cross section of the dark atoms in the present Universe. In the literature, self-interacting dark matter models are usually characterized by their momentum-transfer cross section:
\begin{equation}
\sigma_\text{tr} \equiv \int (1-\cos\theta)\frac{d\sigma_{\rm DM}}{d\Omega} d\Omega,\label{eq:transferdef}
\end{equation}
where $d\sigma_{\rm DM}/d\Omega$ is the differential scattering cross section for dark matter. The angular kernel suppresses irrelevant forward scattering events while giving maximum weight to backward scatterings. For identical particles colliding in the center-of-mass frame (such as two dark atoms), these two processes should be indistinguishable and the above expression overestimates the actual cross section. However, since a dark matter ``particle'' in a typical $N$-body simulation is composed of an extremely large number of actual dark matter particles, there is a fair amount of uncertainty about which microphysical cross section best describes the collisions of these dark matter macro-particles. To facilitate comparison with previous works on SIDM, we choose to parametrize our benchmark models in terms of the commonly accepted $\sigma_\text{tr}$.

Due to the complex internal structure of atoms, their elastic scattering cross section generally has a highly nontrivial energy and angular dependence, with many resonances appearing as the collision energy is varied \cite{Cline:2013pca}. Since our goal in this work is to explore the possible interplay between dark matter self-interaction and the modified matter power spectrum caused by the relativistic force carrier, we consider simplified prescriptions for the dark matter scattering cross section. First, we assume that all of the dark matter is made of neutral dark atoms (that is, we ignore any residual ionization left over after dark recombination) and we also neglect inelastic scattering processes that could excite or even ionize the dark atoms. Such inelastic processes can lead to a very rich phenomenology on sub-galactic scales (see \emph{e.g.} Refs.~\cite{Fan:2013tia,Fan:2013yva,McCullough:2013jma}), but it is beyond the scope of this paper to analyze their consequences in detail. Second, we model neutral dark atom collisions with a hard-sphere scattering process with a constant cross section. While this choice is not necessarily self-consistent within our dark atom model, it allows us to cleanly explore any possible connection between the oscillations and damping in the matter power spectrum and the self-scattering of dark matter inside halos and is acceptable as long as the range of velocities being considered is not too large. Our specific choices for the cross sections are discussed in Section~\ref{sec:SI}.

\subsection{Structure Formation and Evolution up to Decoupling}
Returning for a moment to a general SIDM model, we consider the effect of an additional relativistic particle in the early Universe. We will assume that the dark matter is in thermal equilibrium, and therefore the present-day abundance of dark matter is set either by thermal freeze-out or some asymmetric production mechanism \cite{Kaplan:2009ag,Davoudiasl:2010am,Buckley:2010ui}. Until the light force carrier $\phi$ kinetically decouples or the temperature drops below the carrier mass $m_\phi$, the sound speed in the dark matter bath remains relativistic, and the primordial perturbations in the dark sector undergo collisional (Silk) damping on scales below the dark radiation's diffusion length. The power spectra we will consider are calculated using the  full Boltzmann formalism \cite{Bringmann:2009vf,Cornell:2013rza}, but some intuition can be obtained by considering a simplified version \cite{Hofmann:2001bi}.

Dark matter $D$ is in kinetic equilibrium with the dark radiation $\phi$ until the rate $\Gamma$ of significant energy exchange between the two particle baths is less than the Hubble time $H^{-1}$:
\begin{equation}
\Gamma(T) = n_\phi(T) \langle \sigma v\rangle_{D\phi} v_D^2 \lesssim H(T)^{-1}.  \label{eq:decoupleestimate}
\end{equation}
The extra factors of dark matter velocity $v_D$ are present to account for the multiple scatters that are necessary to significantly change the momentum of a dark matter particle. In dark atom models, the relevant cross sections are those of Eqs.~\eqref{eq:sigmacompton} and \eqref{eq:sigmaraleigh}.

As the Universe expands and cools, primordial perturbations enter the horizon, each characterized by the wavenumber $k$. As long as the dark radiation has not decoupled and the perturbation's wavelength is comparable to the diffusion length of the dark radiation, that mode will be damped, and as a result, no structure of that scale or smaller can be seeded. This will appear as a suppression in the power spectrum of density perturbations of scale $k$. This suppression shares some features with that found in models of WDM. In the latter case, the suppression is due to the high-velocity WDM free-streaming out of the initial overdensities (collisionless damping).

However, unlike the WDM scenario, the matter power spectrum for SIDM with a relativistic force carrier has significant non-trivial structures. These are the result of acoustic oscillations in the dark matter-dark radiation system. On length scales larger than the typical dark radiation mean free path, the dark matter and dark radiation can be considered as a single nearly-perfect fluid. As long as the momentum transfer rate between the dark radiation and the dark matter is large compared to the Hubble rate, the relativistic pressure of the dark fluid leads to a restoring force that effectively opposes the gravitational growth of dark matter overdensities and allow the propagation of longitudinal sound waves in the dark fluid. These sounds waves propagate through the cosmos until the epoch of dark matter kinematic decoupling (estimated by Eq.~\ref{eq:decoupleestimate}) at which point the pressure support falters and dark radiation begins to free-stream out of dark matter fluctuations. Much like the case of their baryonic counterparts, the matter distribution retains a memory of these DAOs which appears as oscillations in the matter power spectrum or as a distinct sound horizon in the matter correlation function.

The specific shape of the SIDM matter power spectrum and correlation function is mostly governed by the relative size of the dark matter sound horizon $r_{\rm DAO}$ and of the diffusion (Silk) damping scale $r_{\rm SD}$ at the epoch of kinematic decoupling. The comoving length-scale is related to the comoving wavenumber $k = \pi/r$. These scales are to a good approximation given by \cite{Cyr-Racine:2013fsa}:

\be\label{eq:r_DAO_theory}
r_{\rm DAO}= \frac{4 \xi^2\sqrt{\Omega_{\gamma}}}{3 H_0 \sqrt{\Omega_{\rm DM}\Omega_{\rm m}}}\ln{\left[\frac{\sqrt{\gamma_{\rm DM}}\sqrt{\Omega_{\rm r}+\Omega_{\rm m} a_D}+\sqrt{\Omega_{\rm m}+\gamma_{\rm DM}a_D}}{\sqrt{\gamma_{\rm DM}\Omega_{\rm r}}+\sqrt{\Omega_{\rm m}}}\right]},
\ee
where we have defined
\be
\gamma_{\rm DM} \equiv \frac{3\Omega_{\rm DM}}{4\xi^4\Omega_{\gamma}};\nonumber
\ee
and
\be\label{eq:r_SD_theory}
r_{\rm SD}\approx \pi\left( \frac{4 a_D^3 m_D}{81 H_0\sqrt{\Omega_{\rm r}} \Omega_{\rm DM}\rho_{\rm crit} \sigma_{\rm Compton}}\right)^{1/2}.
\ee
Here, $a_D$ stands for the scale factor at the epoch of dark matter kinematic decoupling, and $H_0$ is the present-day Hubble constant. $\Omega_{\rm DM}$, $\Omega_{\gamma}$, $\Omega_{\rm r}$, and $\Omega_{\rm m}$ stand for the energy density in dark matter, photons, radiation (including neutrinos and dark radiation), and non-relativistic matter, respectively; all in units of the critical density of the Universe, $\rho_{\rm crit}$. We note that both Eqs.~(\ref{eq:r_DAO_theory})  and (\ref{eq:r_SD_theory}) were derived in the tightly-coupled regime of the dark plasma which is valid until close to the epoch of kinematic decoupling. We immediately see that the dark matter sound horizon and its Silk damping scale depend on different combinations of the cosmological and SIDM parameters, implying that these two scales could be somewhat independently varied. We can thus identify two kinds of models:\footnote{We note that the case $r_{\rm SD} > r_{\rm DAO}$ is ill-defined since this requires the dark radiation to be effectively decoupled from dark matter, in which case no sound wave can propagate in the dark medium. }
\begin{itemize}

\item $r_{\rm SD} \ll r_{\rm DAO}$: In this case, diffusion damping is ineffective on scales close to the dark matter sound horizon. This results in a sharp and localized sound horizon that is imprinted in the dark matter density field after kinematic decoupling. In terms of the matter power spectrum, these models display a significant number of nearly undamped acoustic oscillations before Silk damping becomes effective and dramatically reduces power on small length scales. Generically, this category encompasses models with large values of $\sigma_{\rm Compton}/m_D$.

\item $r_{\rm SD}\sim r_{\rm DAO}$: For these models, the dark radiation diffusion scale is comparable to the dark matter sound horizon at decoupling, leading to a substantial damping of the acoustic oscillations in the dark plasma. This diffusion damping significantly broadens and dilutes the dark matter sound horizon, leaving only the small-scale suppression of structure as the key signature of these models. In this case, we expect the matter power spectrum to display only a handful of strongly damped oscillations.

\end{itemize}
Since models falling into the first category (which we will call ``strong'' DAO) are characterized by two distinct scales, they contrast significantly with WDM theories whose cosmological behavior is uniquely determined by their free-streaming length. On the other hand, the cosmological observables of models falling into the second category (``weak'' DAO) are mostly determined by one scale, the Silk damping length, implying that these scenarios might be harder to distinguish from a WDM model once nonlinear evolution is taken into account.  That said, the dark matter self-interactions will generically lead to a different internal structure for dark matter halos (as we will discuss next), giving us another handle to distinguish the SIDM scenarios from WDM models. The $N$-body simulations presented in the next few sections aim at determining whether such distinction is possible and also whether the sound horizon and and Silk damping scale remain separately imprinted in the nonlinear matter field.

\begin{figure*}[t!]
\subfigure{\includegraphics[width=0.49\textwidth]{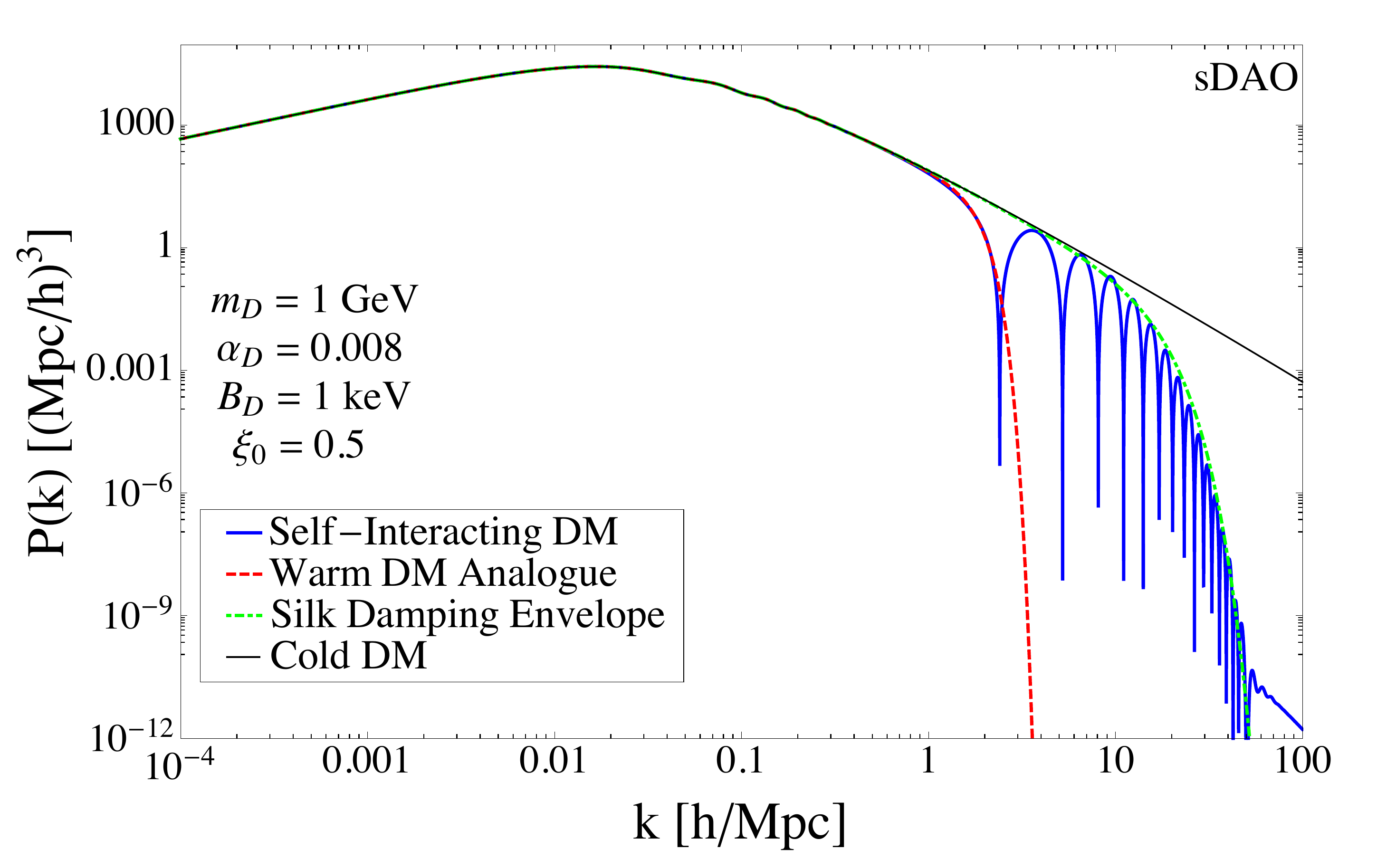}}
\subfigure{\includegraphics[width=0.49\textwidth]{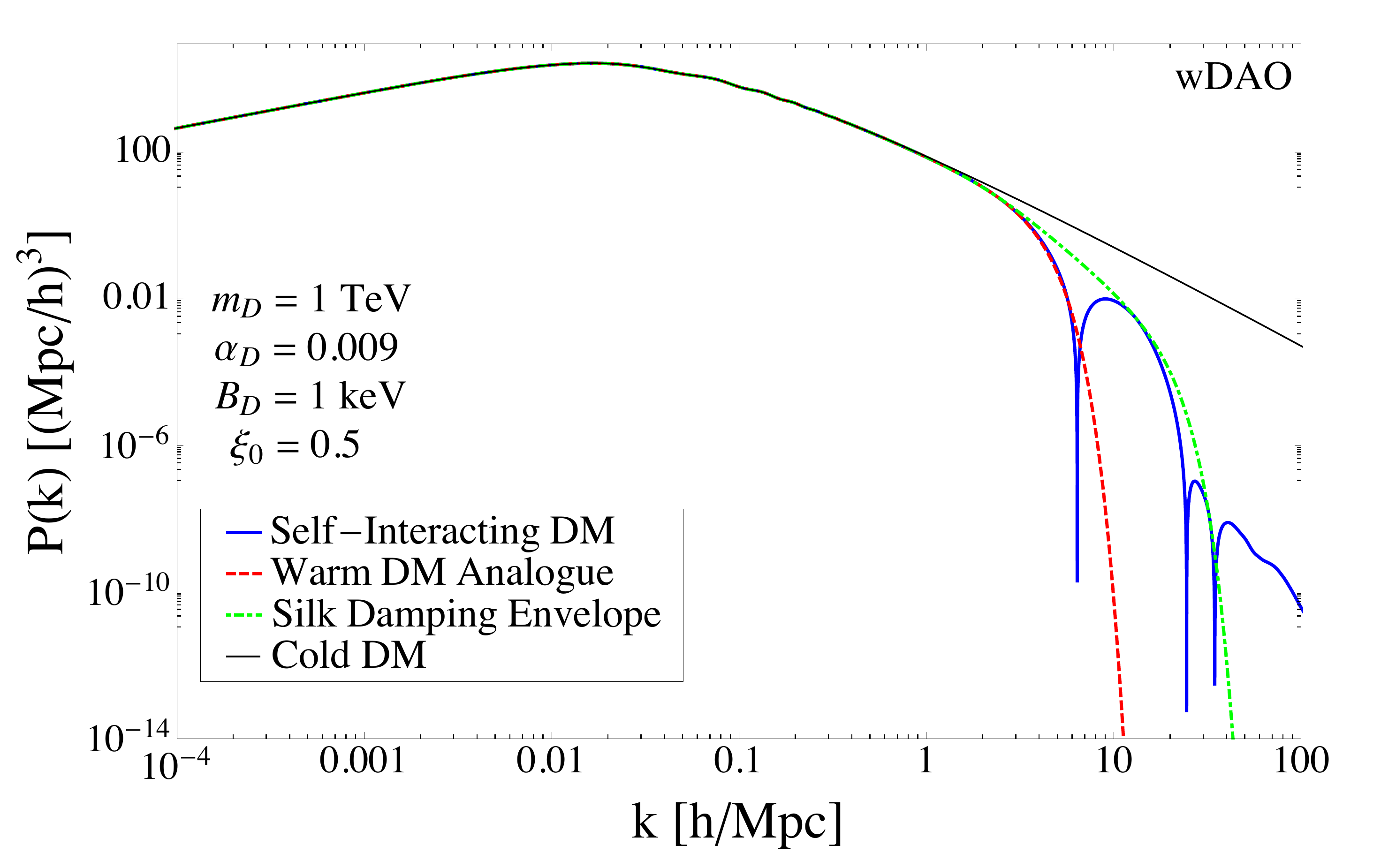}}
\caption{Comparison between the linear matter power spectra as a function of wavenumber $k$ for SIDM with a light mediator (here, dark atoms) and that of WDM with a free-streaming length comparable to the sound horizon of the former. We also display the standard matter power spectrum for cold collisionless dark matter as well as a fit to the Silk damping envelope of SIDM. The left panel displays the benchmark model for which $r_{\rm DAO} \gg r_{\rm SD}$ (strong DAO), while the right panel shows the scenario for which $r_{\rm DAO} \sim r_{\rm SD}$ (weak DAO). Here, $\xi_0\equiv\xi(T_{\rm CMB,0})$.  }
\label{fig:examplepowerspectrum}
\end{figure*}

\begin{figure*}
\begin{centering}
\subfigure{\includegraphics[width=0.49\textwidth]{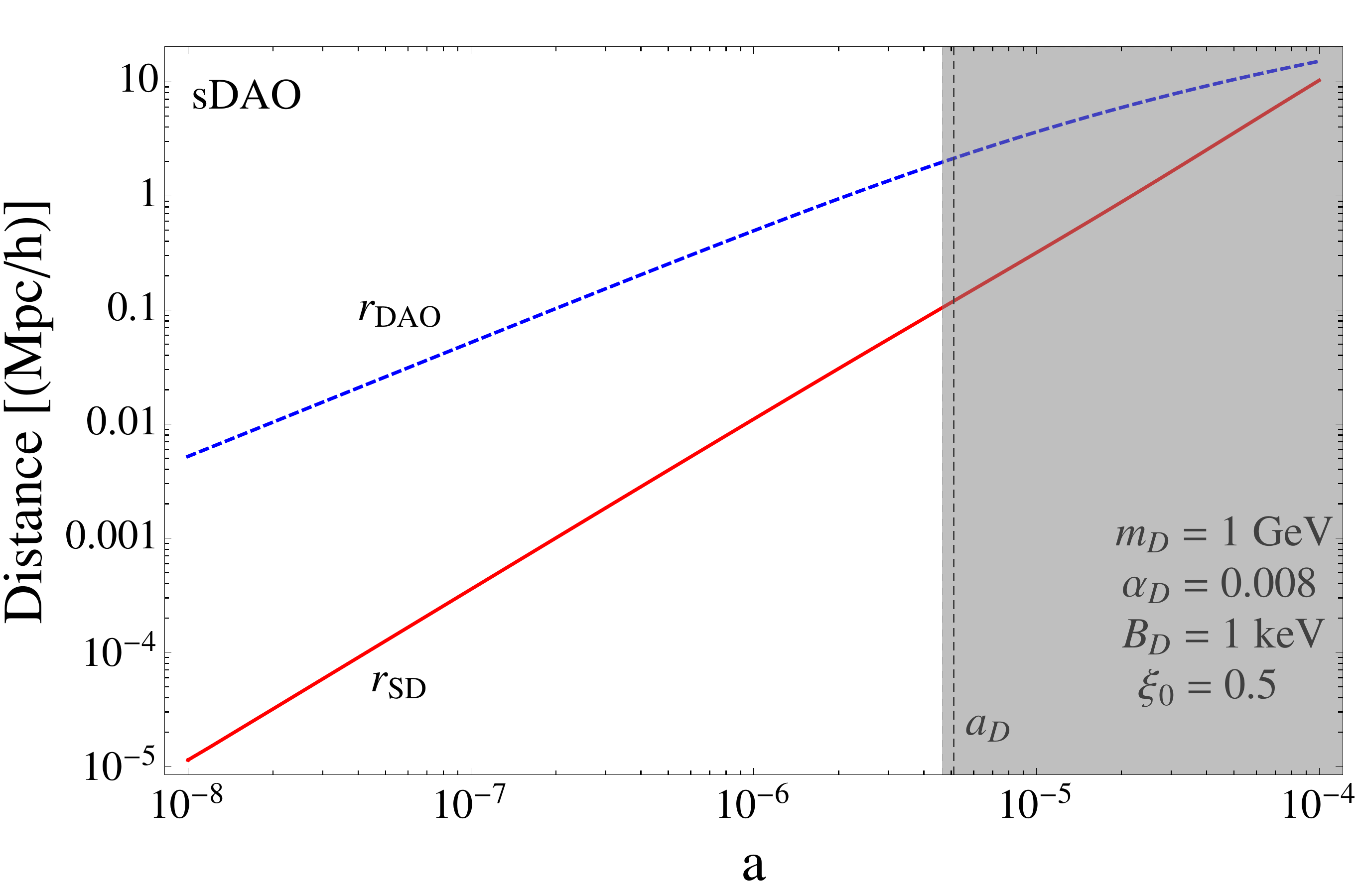}}
\subfigure{\includegraphics[width=0.49\textwidth]{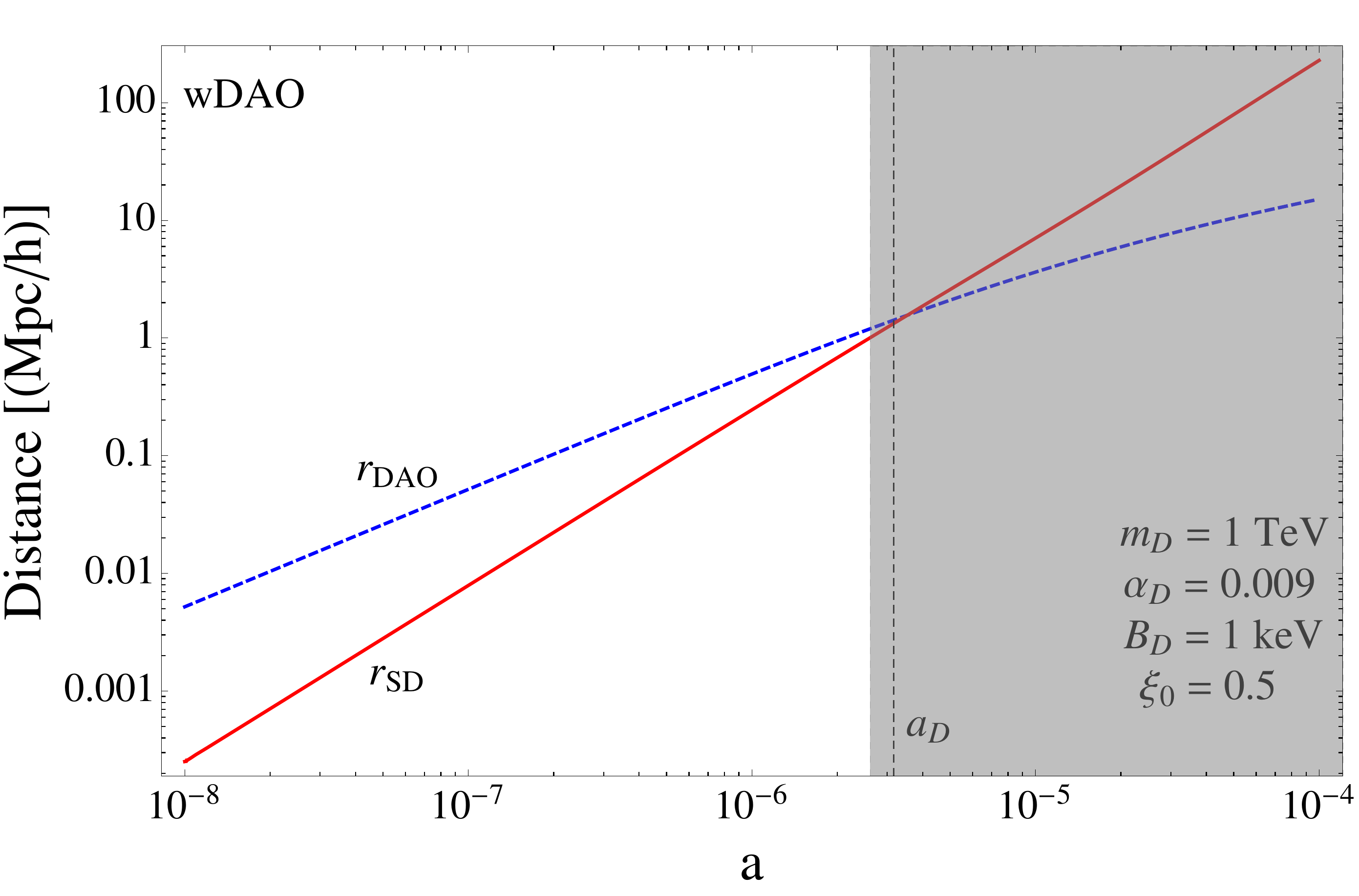}}
\caption{Evolution of the size of the dark matter sound horizon ($r_{\rm DAO}$) and of the Silk damping scale ($r_{\rm SD}$) as a function of the cosmological scale factor $a$. The vertical dashed line denotes the epoch of kinematic decoupling. The grayed regions denote where our calculation of these scales breaks down.  The left panel displays the strong DAO model for which $r_{\rm DAO} \gg r_{\rm SD}$ always, while the right panel shows the weak DAO model with $r_{\rm DAO} \sim r_{\rm SD}$ at $a=a_D$. Here, $\xi_0\equiv\xi(T_{\rm CMB,0})$.  }
\label{fig:rDAOvsrSD}
\end{centering}
\end{figure*}

In Fig.~\ref{fig:examplepowerspectrum}, we show the linear power spectrum of CDM, compared to that of a dark atom model, with two benchmark parameter sets that exemplify strong (left panel) and weak (right panel) DAOs. The power spectrum is calculated using the full Boltzmann equations for dark matter coupled to dark radiation \cite{CyrRacine:2012fz}. The two parameters sets are:
\begin{eqnarray}
\mbox{{\bf Strong DAO:}} & & m_D=1~\mbox{GeV},~\alpha_D = 8 \times 10^{-3},~B_D = 1~\mbox{keV},~\xi(T_{\rm CMB,0}) = 0.5 \label{eq:strongDAOparam} \\
\mbox{{\bf Weak DAO:}} & & m_D=1~\mbox{TeV},~\alpha_D = 9 \times 10^{-3},~B_D = 1~\mbox{keV},~\xi(T_{\rm CMB,0}) = 0.5, \label{eq:weakDAOparam} 
\end{eqnarray}
where $T_{\rm CMB,0}$ is the temperature of the CMB today. In this paper, we will denote the two models as ADM$_\text{sDAO}$ and ADM$_\text{wDAO}$.  We note that both models considered in this work are in agreement with the cosmological constraints presented in Ref.~\cite{Cyr-Racine:2013fsa}. In the ADM$_\text{sDAO}$ case, we observe that the power spectrum displays a number of nearly-undamped oscillations before the Silk damping cutoff  (dot-dashed damping envelope) becomes important on smaller scales. In contrast,  for the ADM$_\text{wDAO}$ case even the first oscillation is strongly Silk-damped as compared to the CDM amplitude. In both cases, we observe that the overall shape of the linear matter power spectrum of SIDM models with long range forces significantly departs from that of WDM and CDM (also shown in Fig.~\ref{fig:examplepowerspectrum}) on small length scales, but is otherwise identical to CDM on larger cosmological scales. The evolution of the two key scales, $r_\text{SD}$ and $r_\text{DAO}$, as a function of the scale factor $a$ is shown in Fig.~\ref{fig:rDAOvsrSD}. The scale factors of kinetic decoupling $a_D$, used in Eqs.~\eqref{eq:r_DAO_theory} and \eqref{eq:r_SD_theory}, are also shown as vertical dashed lines. As expected, $(r_{\rm DAO}/r_{\rm SD})|_{a=a_D}\gg1$ in the strong DAO case, while $(r_{\rm DAO}/r_{\rm SD})|_{a=a_D}\sim1$ in the weak DAO case.


In this work, we are interested in the impact of the dark matter microphysics (through its effect on the matter power spectrum and the self-scattering cross section) on the number density and distribution of small scale structure in the Universe. It is therefore useful to convert the length scales $r_\text{DAO}$ and $r_\text{SD}$ (or, their equivalent wavenumbers) into the mass of a collapsed dark matter halo of the corresponding size. The mass of dark matter enclosed today by wavenumber $k$ is approximately:
\begin{equation}\label{eq:Mcut}
M(k) \approx (10^{12}~M_\odot) \left(\frac{k}{\mbox{Mpc}^{-1}}\right)^{-3}.
\end{equation}

For comparison, in supersymmetric models with a ``standard'' neutralino dark matter candidate, the mass cut-off in the power spectrum is set by the temperature at which the dark matter kinetically decouples from the relativistic Standard Model neutrinos. Under reasonable assumptions for the neutralino physics, this occurs around $T \sim 30$~MeV \cite{Hofmann:2001bi,Chen:2001jz,Green:2003un,Loeb:2005pm}. The physical Jeans wavenumber, setting the scale at which perturbations will begin gravitational collapse (assuming sound speed $v_s$) is:
\begin{equation}
k_J = \left( \frac{4\pi\rho(T)}{m_\text{Pl}^2 v_s^2}\right)^{1/2}.
\end{equation}
Here $\rho(T)$ is the total energy density of the Universe at temperature $T$. Assuming that the Universe is radiation-dominated at this point in its history, the Jeans wavenumber for such models is $k_J \sim 10^6$~Mpc$^{-1}$, and so dark matter halo masses extend down to $\lesssim 10^{-6}M_\odot$. In the dark atom SIDM model with self-scattering cross sections large enough to effect the evolution of galactic structures, the decoupling temperature (or mediator mass) is expected to be much lower, in the range of keV or tens of eV, resulting in a suppression of structure on the scale of dwarf galaxies or larger. As can be seen in Fig.~\ref{fig:examplepowerspectrum}, the characteristic scale at which the power spectrum deviates from that of CDM is 1-10~$h/$Mpc$^{-1}$ for our two benchmark scenarios. From the previous arguments, for the dark atom models under consideration here, we expect suppression of dark matter structure to begin at scales between $10^{9}$ and $10^{12}M_\odot$, though we point out that lower end of this mass range will be below the resolution limit of our simulations.

The scales over which these ADM models deviate from the predictions of CDM are constrained primarily by measurements of the Lyman-$\alpha$ forest \cite{McDonald:2004eu,Seljak:2004xh,Tegmark:2002cy}, though uncertainties exist in the conversion between the primordial power spectrum and the observations \cite{Bird:2010mp}. As pointed out in Ref.~\cite{CyrRacine:2012fz}, the addition of ADM would require hydrodynamical simulations of the ADM cosmology to accurately apply the Lyman-$\alpha$ constraints. With this caveat in mind, the Lyman-$\alpha$ data sees no deviation from $\Lambda$CDM on scales larger than $k < 2h$~Mpc$^{-1}$ \cite{Seljak:2006qw} or $5 h$~Mpc$^{-1}$ \cite{Croft:2000hs,Gnedin:2001wg,Tegmark:2002cy}. This corresponds to a minimum halo mass of $10^{11-13}M_\odot$. The lower edge of this mass range would be in tension with our ADM power spectra, though again, care must be taken in directly extrapolating the bounds to scenarios with new dark matter physics.

\subsection{Effects of Self-Interactions on Post-Decoupling Halo Evolution}\label{sec:SI}
As covered in the previous two sub-sections, the presence of a light particle coupled to the dark matter in the early Universe impacts the matter power spectrum, imprinting it with both DAOs and a damping scale. Once the dark matter decouples from the dark radiation, the interaction mediated by the light particle causes the dark matter to evolve forward as a self-interacting cold dark matter particle (albeit one with modified initial conditions). If the self-interaction is sufficiently large, it will effect energy and momentum transfer in dark matter halos, transforming central density cusps into cores \cite{Spergel:1999mh}, changing velocity distributions and smoothing velocity profiles \cite{Vogelsberger:2012ku,Rocha:2012jg,Vogelsberger:2013}, and reducing the triaxiality of galaxies \cite{Peter:2012jh}. Some have claimed that SIDM models can bring simulation and observation into closer alignment  \cite{Spergel:1999mh,Vogelsberger:2012ku,Rocha:2012jg,Zavala:2012us}, though this is a subject of active debate. We are interested in the interplay of the SIDM scattering effects with the alteration of the initial matter power spectrum.

The figure of merit for SIDM models is the momentum-transfer cross section per dark matter mass, $\sigma_\text{tr}/m_D$, where  $\sigma_\text{tr}$ is given in Eq.~\eqref{eq:transferdef}. As a general rule, any ``realistic'' model of SIDM will have $\sigma_\text{tr}$ be a non-trivial function of the relative velocity $v$. For example, scattering of particles charged under an unbroken $U(1)$ gauge field is proportional to $v^{-4}$. Models with a massive mediator have a more complicated velocity dependence. If the spectrum admits bound states, resonances can develop, leading to $\sigma_\text{tr}$ varying greatly as a function of $v$ \cite{Buckley:2009in,Loeb:2010gj,Tulin:2012wi,Tulin:2013teo,Cline:2013pca}. This is equivalent to the Sommerfeld enhancement which is possible in dark matter annihilation \cite{ArkaniHamed:2008qn}.

As these examples indicate, it is not uncommon for a model of SIDM to have transfer cross sections with a complicated dependence on $v$. As discussed in Section~\ref{sec:darkatoms}, our benchmark dark atom model certainly has such a non-trivial functional form, with many molecular resonances appearing as the collision energy is varied. As our interest in this paper is merely demonstrating the general properties of SIDM models coupled to a light mediator, our simulation suite does not have the sufficient resolution that would require a more detailed treatment of the $v$-dependence of $\sigma_\text{tr}$.  

Very roughly, observations exclude values of $\sigma_\text{tr}/m_D$ large enough to cause approximately one scattering  per particle per dynamical time of the relevant system. From observations of the Bullet Cluster, $\sigma_\text{tr}/m_D < 1.25$~cm$^2$/g for velocities on the order of 1000~km/s \cite{Clowe:2006eq,Randall:2007ph}. The shape of galaxy clusters and massive elliptical galaxies also indicate that $\sigma_\text{tr}/m_D \lesssim 1$~cm$^2$/g for similar characteristic velocities \cite{Rocha:2012jg}. To create observable cored profiles for dwarf galaxies ($v \sim 30-60$~km/s), $\sigma_\text{tr}/m_D$ must be larger than 0.1~cm$^2$/g \cite{Loeb:2010gj}, while a minimum of 0.6~cm$^2$/g is needed to reduce the central masses these dwarfs \cite{Zavala:2012us} sufficiently to solve the ``Too Big to Fail'' problem \cite{BoylanKolchin:2011de} 

For our weak DAO benchmark model, analytical estimates of the dark matter transfer cross section over mass (see \emph{e.g.}~Ref.~\cite{Cline:2013pca}) show that it is negligibly small for the typical velocities of interest ($v\sim100-1000$ km/s) that we are able to resolve with our simulations. For instance, $(\sigma_\text{tr}/m_D)_\text{weak}\sim0.1$ cm$^2$/g at $v=220$ km/s while it drops to $(\sigma_\text{tr}/m_D)_\text{weak}\sim5\times10^{-3}$ cm$^2$/g at $v=500$ km/s, where $v$ is the relative velocity of colliding particles. Such a small cross section over mass is mainly caused by the large dark atom mass in this model ($m_D=1$ TeV) which suppresses the number density of dark matter particles. We therefore neglect\footnote{We have explicitly verified that, on the mass scales probed by our simulations, the collisions have negligible effects for the weak DAO case.} collisions between dark matter particles for the weak DAO case, but we emphasize that they would become important in higher resolution simulations. For the strong DAO benchmark case, we have the opposite problem that the transfer cross section over mass is much larger than the bounds mentioned above at the velocities of interest. To avoid simulating a grossly unrealistic case, we instead assign to this model a constant value for the transfer cross section over mass that is in the interesting range to potentially address the small-scale astrophysical problems:
\begin{eqnarray}
(\sigma_\text{tr}/m_D)_\text{strong} & = & 1 \mbox{ cm$^2$/g}.
\end{eqnarray}

We note that despite our apparent \emph{ad hoc} choices above, there are a large number of dark atom models that naturally have $\sigma_\text{tr}/m_D\sim 0.1-1$ cm$^2$/g at velocities relevant to astrophysical objects such as dwarf galaxies and clusters. However, these models typically have Silk damping mass scales $M_{\rm SD} < 10^9 M_{\odot}$ (see Eq.~\ref{eq:Mcut}), which would be much more computationally expensive. Nevertheless, we expect the general conclusions drawn from our current $N$-body simulations to be broadly applicable to fully realistic dark atom simulations as well as to other SIDM models coupled to a light force carrier, and future work at higher resolution could probe this interesting range of dark atom parameters.
\section{Simulation of Dark Atoms \label{sec:simulation}}
Having introduced the three handles ($r_\text{DAO}$, $r_\text{SD}$, and $\sigma_\text{tr}/m_D$) that together can distinguish a SIDM model with a long-range force from CDM, WDM, or ``regular'' SIDM models, we wish to investigate the observable differences between these sets of models. To do so, we use fully cosmological $N$-body simulations of dark matter to compare the properties of dark matter halos formed in each scenario, which emphasize different phenomenological aspects of long-range SIDM.

The two dark atom models we consider are given by the strong and weak DAO parameter sets given by Eqs.~\eqref{eq:strongDAOparam} and \eqref{eq:weakDAOparam}. As can be seen in Fig.~\ref{fig:rDAOvsrSD}, the value of $r_\text{DAO}$ for these two models is similar. To demonstrate the importance of the two scales $r_\text{DAO}$ and $r_\text{SD}$ inherent in the ADM models, we compare to an additional two sets of models which have power spectra suppressed at a single scale when compared to the CDM scenario. First, a WDM model is used, chosen to have a free-streaming damping scale equivalent to the $r_\text{DAO}$ of ADM$_\text{sDAO}$ (which is also close to the sound horizon of ADM$_\text{wDAO}$). Next, a modified ADM model is constructed, the power spectrum of which is fit to the damping envelope of the ADM$_\text{sDAO}$ model. This model is somewhat {\it ad hoc}, as -- lacking acoustic oscillations --  it is not a model of atomic dark matter that can be realized with any choice of parameters. However, it isolates the effect of the single $r_\text{SD}$ scale, and so is useful for comparison purposes. 

In addition, for the strong DAO model, we run simulations both with and without the self-interactions between dark matter particles. This allows us to isolate the effect of collisions on the structure and number of dark matter halos. The non-collisional simulations will be denoted by the subscript ``nc.'' In the weak DAO case, only non-collisional runs are performed since the relevant $\sigma_\text{tr}/m_D$ is too small to have significant effects at the resolved scales of the simulation, as mentioned in the previous section. Thus, for ADM$_\text{wDAO}$, only the effects of the modifications to the primordial power spectrum are probed in this work, though future simulations at higher resolution could resolves effects of the non-zero transfer cross section. 

For each of these models, we choose a flat cosmology that is consistent\footnote{Taking into account the extra dark radiation in our model.} with the results from the  Planck mission \cite{planckXVI} with a corresponding linear power spectrum (at $z=0$) as shown in Fig.~\ref{fig:examplepowerspectrum}. These cases have the following cosmological parameters (of relevance for the simulations): $\Omega_{\rm m}=0.305$, $\Omega_\Lambda=0.695$, $H_0=100h$km~s$^{-1}$~Mpc$^{-1}$ with $h=0.696$, $n_{\rm s}=0.97$ and $\sigma_8=0.86$, where $n_{\rm s}$ is the spectral index of the primordial power spectrum, and $\sigma_8$ is the rms amplitude of linear mass fluctuations in $8~h^{-1}\,{\rm Mpc}$ spheres at redshift zero. 

The simulations follow the growth of dark matter structure from $z=127$ to $z=0$ in a cubic box of size $L=64h^{-1}$~Mpc with $512^3$ simulation particles starting from the same initial conditions (save for the varying power spectra across models) with a fixed comoving softening length (Plummer-equivalent), $\epsilon\sim2.8h^{-1}$~kpc. The dark matter particle mass is $m_{\rm p}\sim1.65\times10^8h^{-1}$~M$_\odot$ and the Nyquist frequency is $\sim25h$~Mpc$^{-1}$. For the ADM simulations, an algorithm for elastic isotropic self-scattering is implemented on top of the $N-$body code {\tt GADGET-3} for gravitational interactions (last described in Ref.~\cite{Springel:2005mi}) as described in detail in Ref.~\cite{Vogelsberger:2012ku}. The algorithm is based on a Monte Carlo approach to represent the microphysical scattering process in the macroscopic context of the simulation. The main properties of our simulation suite are given in Table \ref{table:simulations}. The last column of this table gives an exponential suppression scale, which is set by the collisionless damping scale for WDM, $r_\text{SD}$ for ADM$_\text{sDAO}$ and ADM$_\text{wDAO}$, and the fitted envelope (effectively $r_\text{SD}$) for ADM$_\text{sDAO-env}$. Using convergence tests, we show in the Appendix that the effective halo mass resolution for our simulations is better than $10^{11}~h^{-1}$M$_\odot$ and that inner densities of dark matter halos can be trusted at radius of $\sim 3\epsilon$, or $\sim 8.4h^{-1}$kpc.

\begin{table}
\begin{tabular}{lccc}
\hline
name                   & $\sigma_{\rm tr}/m_D$ [cm$^2$/g]\,\,    & $(r_{\rm DAO}/r_{\rm SD})|_{a=a_D}$\,\,    & Suppression Scale [Mpc/h]\\
\hline
\hline
CDM               & $--$            & $--$                & $--$           \\
WDM             & $--$           & $--$                & $2.1$\footnote{Chosen such that it matches the DAO scale of ADM$_{\rm sDAO}$.}         \\
ADM$_{\rm wDAO}$(nc)            & $0.1$\footnote{Actual value of $\sigma_{\rm tr}/m_D$ evaluated $v=220$ km/s, but not used in our simulations since it is too small to be relevant.}            & $1.07$                & $1.5$           \\
ADM$_{\rm sDAO-env}$(nc)           & $--$           & $--$               & $0.12$          \\
ADM$_{\rm sDAO}$           & $1.0$\footnote{Independent of velocity.}            &     $17.9$           & $0.12$           \\
ADM$_{\rm sDAO}$(nc)           & $--$            & $17.9$             & $0.12$        \\

\hline
\end{tabular}
\caption{Simulation suite discussed in this work. Only one simulation has self-scattering (ADM$_{\rm sDAO}$), the rest are collisionless. The second column characterizes the relevance of dark acoustic oscillations to the Silk damping scale. The third column gives an estimate of the scale at which the non-CDM models' power spectra is exponentially suppressed as compared to the CDM benchmark. This is the collisionless damping scale for the WDM model, and the $r_\text{SD}$ scale for the ADM and ADM-derived models.}
\label{table:simulations}
\end{table}
\section{Effects of acoustic oscillations and the damping scale \label{sec:dao}}
\begin{figure}[t]
\includegraphics[width=0.49\columnwidth]{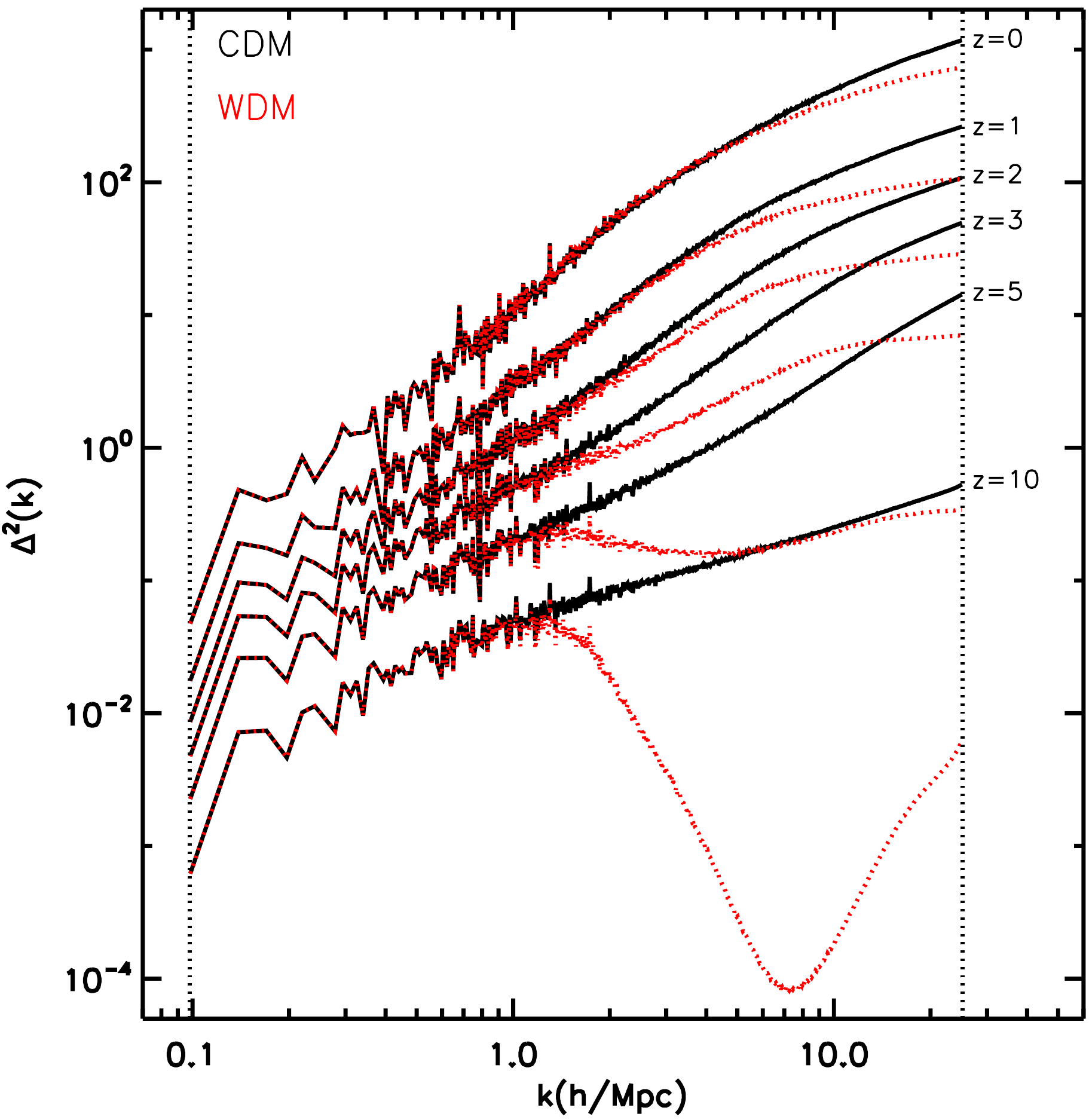}
\includegraphics[width=0.49\columnwidth]{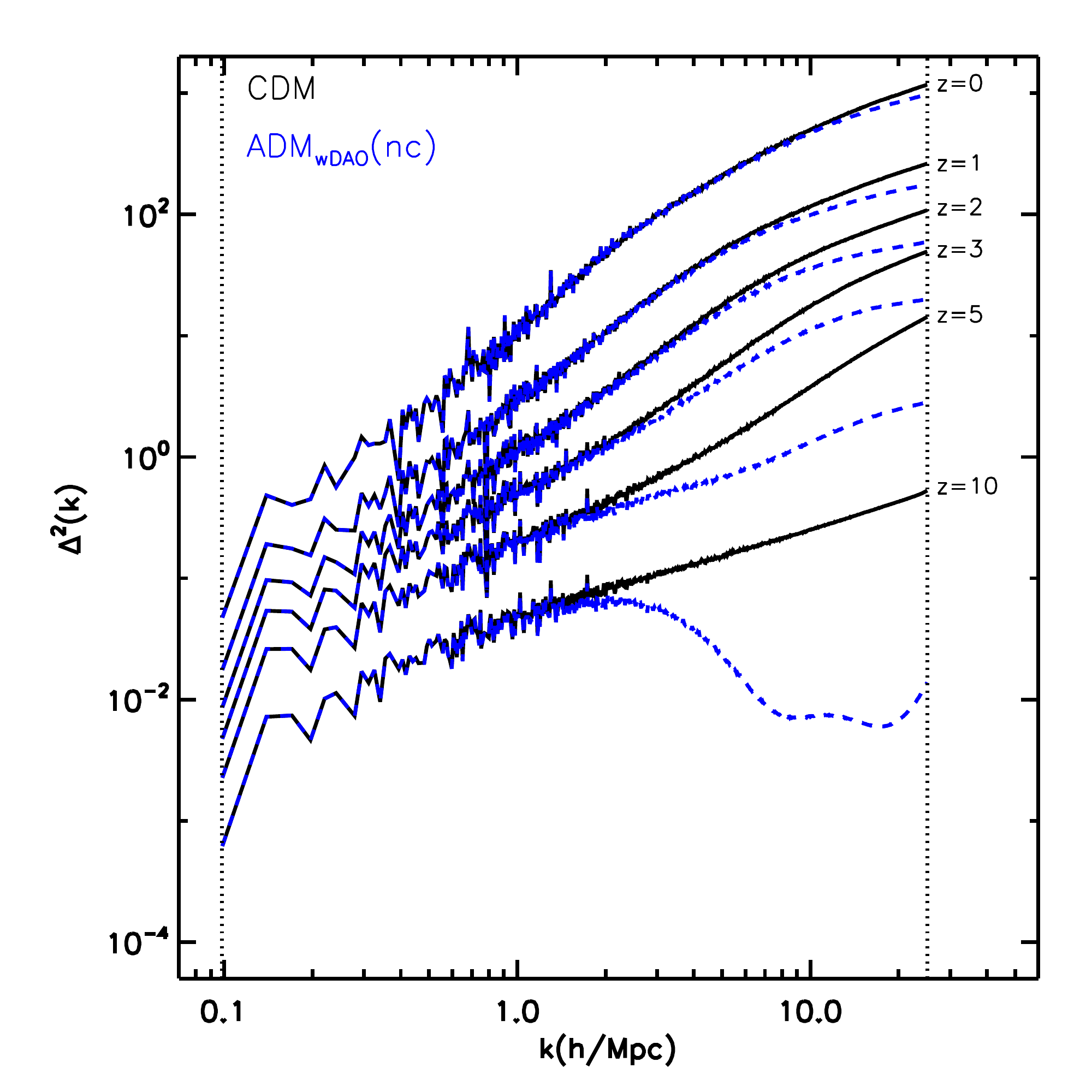}
\includegraphics[width=0.49\columnwidth]{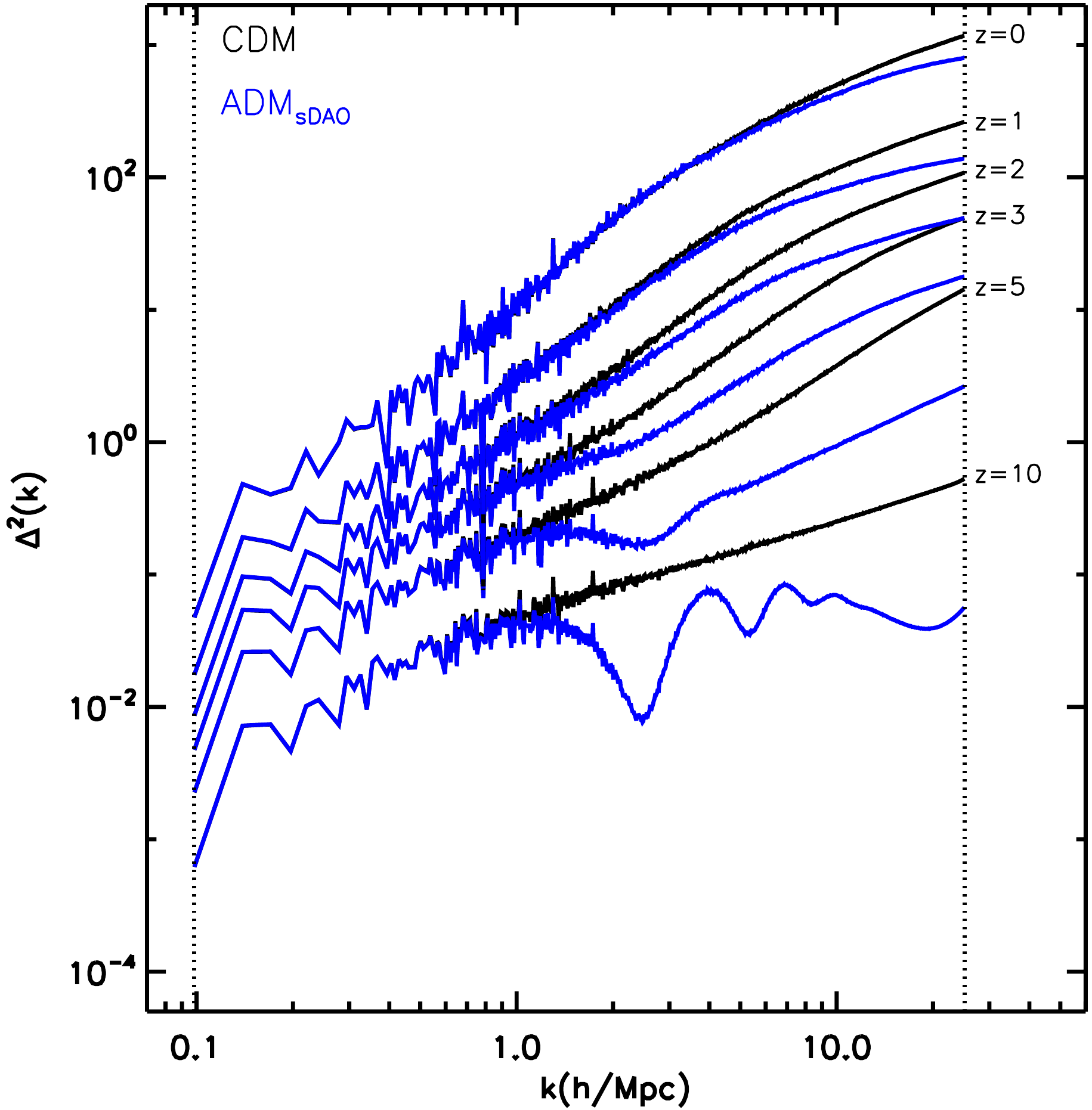}
\caption{Evolution of the dimensionless matter power spectrum $\Delta^2(k)\equiv k^3P(k)$ for the WDM (upper left, dotted red line), ADM$_\text{wDAO}$ (upper right, dashed blue line), and ADM$_\text{sDAO}$ (lower panel, solid blue line) simulations. In each case, the CDM power spectrum is shown in black for comparison. We display the fully nonlinear $\Delta^2(k)$ evaluated at 6 different redshifts from $z=10$ to $z=0$. The dotted vertical lines at large and small $k$ values denote the Nyquist frequency of our simulation box and the largest scale (fundamental mode) probed by our simulations, respectively.}
\label{fig:power_spectrum_evol}
\end{figure}
Using our simulations, we can now investigate the nonlinear evolution of the matter power spectrum in SIDM models with a light force carrier. Given the resolved scales of these simulations, and the relatively low, velocity-independent self-scattering cross section employed, we do not expect to see significant deviations in the number density of halos between the collisional and non-collisional simulations. As result, this section is most devoted to the observable results of the initial matter power spectrum being imprinted with the scales $r_\text{SD}$ and $r_\text{DAO}$. At the resolution of these simulations, non-zero self-interaction cross sections can alter the inner structure of low-mass halos, which we will explore in Section~\ref{sec:density}.

Fig.~\ref{fig:power_spectrum_evol} shows the redshift evolution of the nonlinear matter power spectrum for three of our non-CDM simulations, in each case comparing it to that of CDM. We display dimensionless nonlinear power spectra ($\Delta^2(k)\equiv k^3P(k)$) evaluated at six different redshifts ranging from $z=10$ to $z=0$. For both the strong and weak DAO cases, we observe that the non-linear evolution progressively erases the acoustic oscillations and regenerates power on scales initially affected by DAOs. At redshift zero, the nonlinear matter power spectrum of our ADM models closely resemble that of CDM, except for a modest suppression on scales $k\gtrsim 5h$ Mpc$^{-1}$ (in the regime dominated by correlations among dark matter particles within individual halos, the {\it 1-halo} term). A suppression is also observed for our WDM benchmark at $z=0$ for large wave numbers, but its magnitude is slightly larger than in the ADM$_{\rm sDAO}$ case, in line with our expectations given the absence of acoustic oscillations in this model. At first sight, it thus seems that nonlinearities erase the distinction between $r_{\rm DAO}$ and $r_{\rm SD}$ in $\Delta^2(k)$ at low redshifts in SIDM models with relativistic force carriers, effectively replacing these two quantities by a single effective damping scale. However, as we discuss below, the actual situation is more subtle and interesting.
\begin{figure}[t]
\includegraphics[width=0.45\columnwidth]{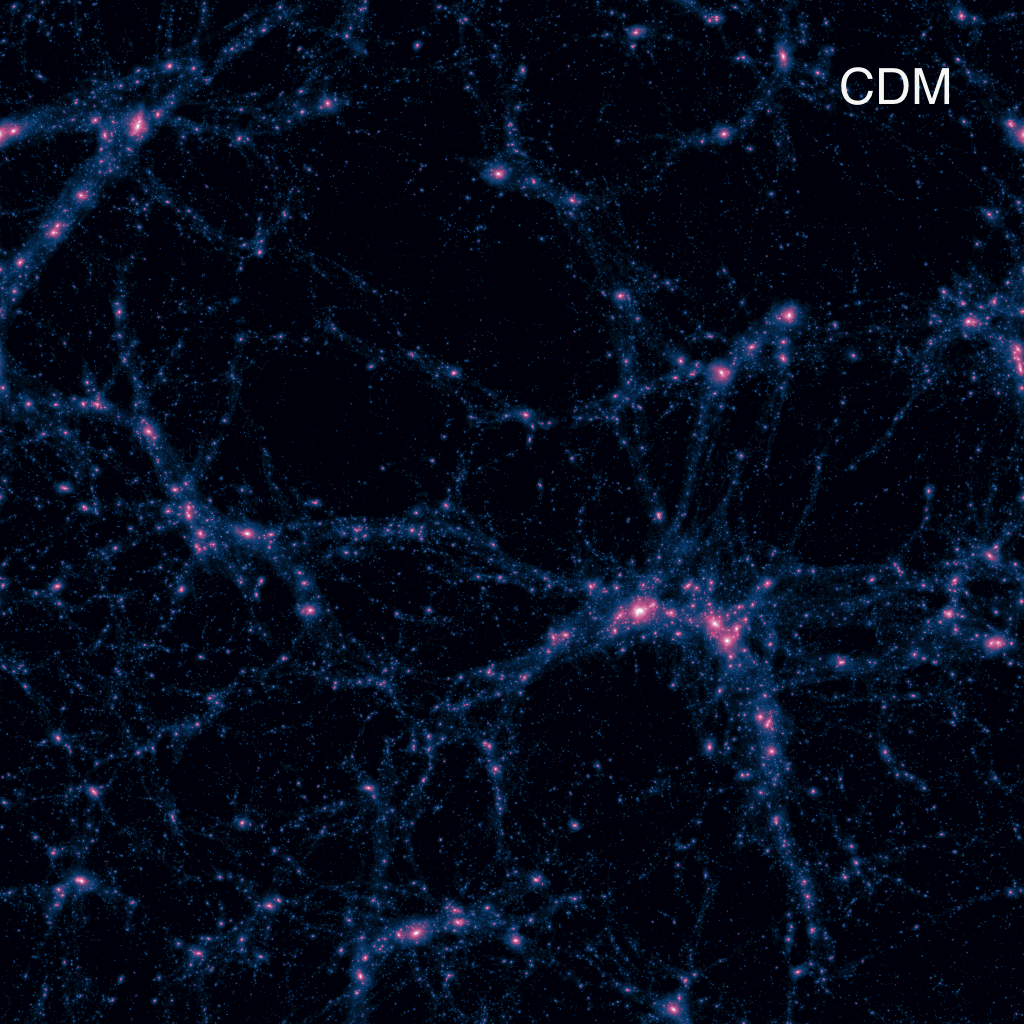}
\includegraphics[width=0.45\columnwidth]{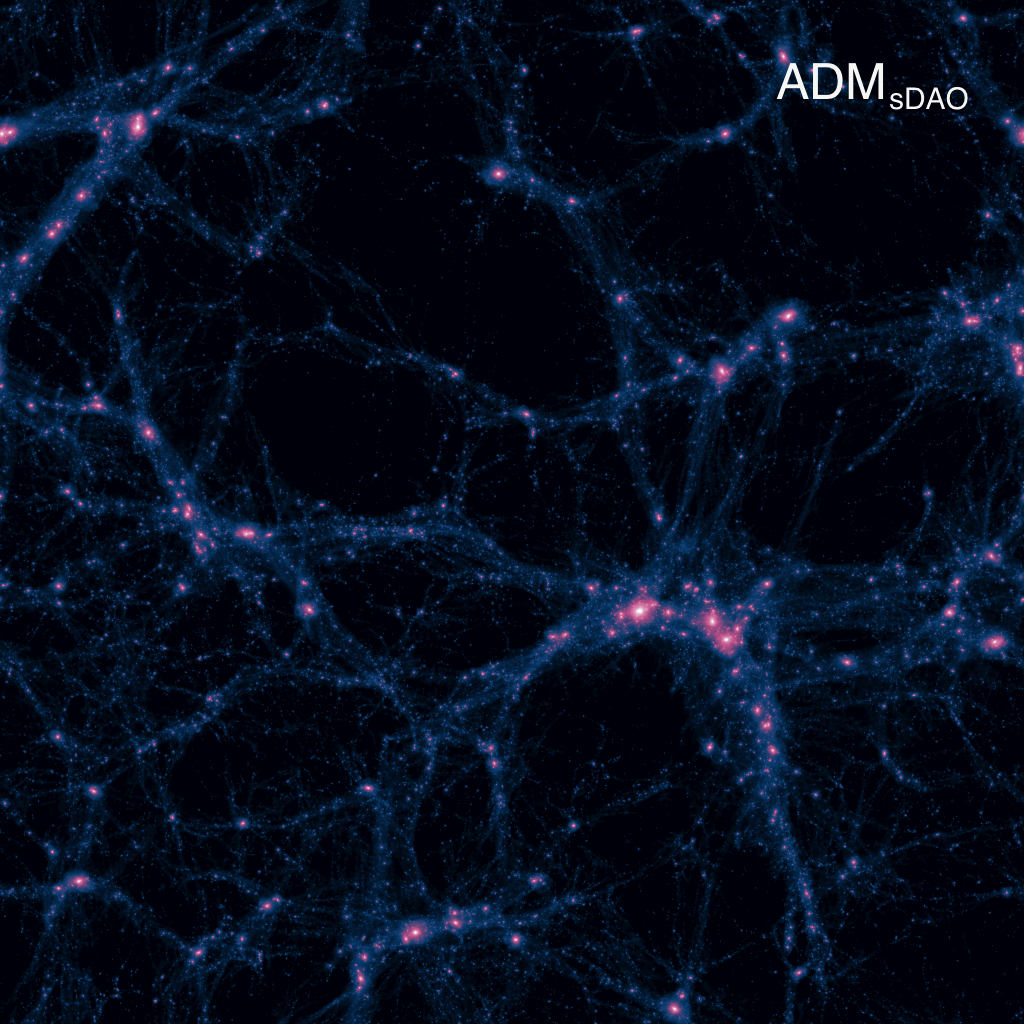}
\includegraphics[width=0.45\columnwidth]{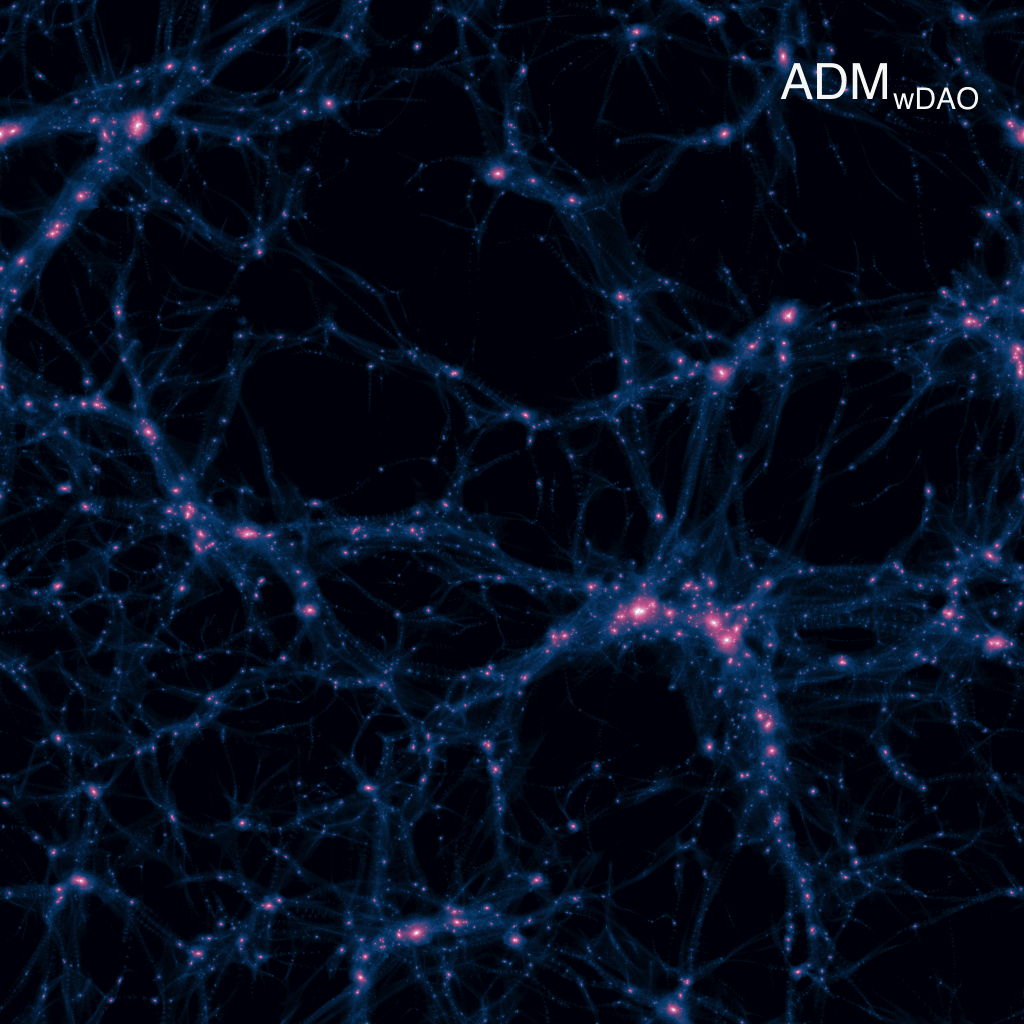}
\includegraphics[width=0.45\columnwidth]{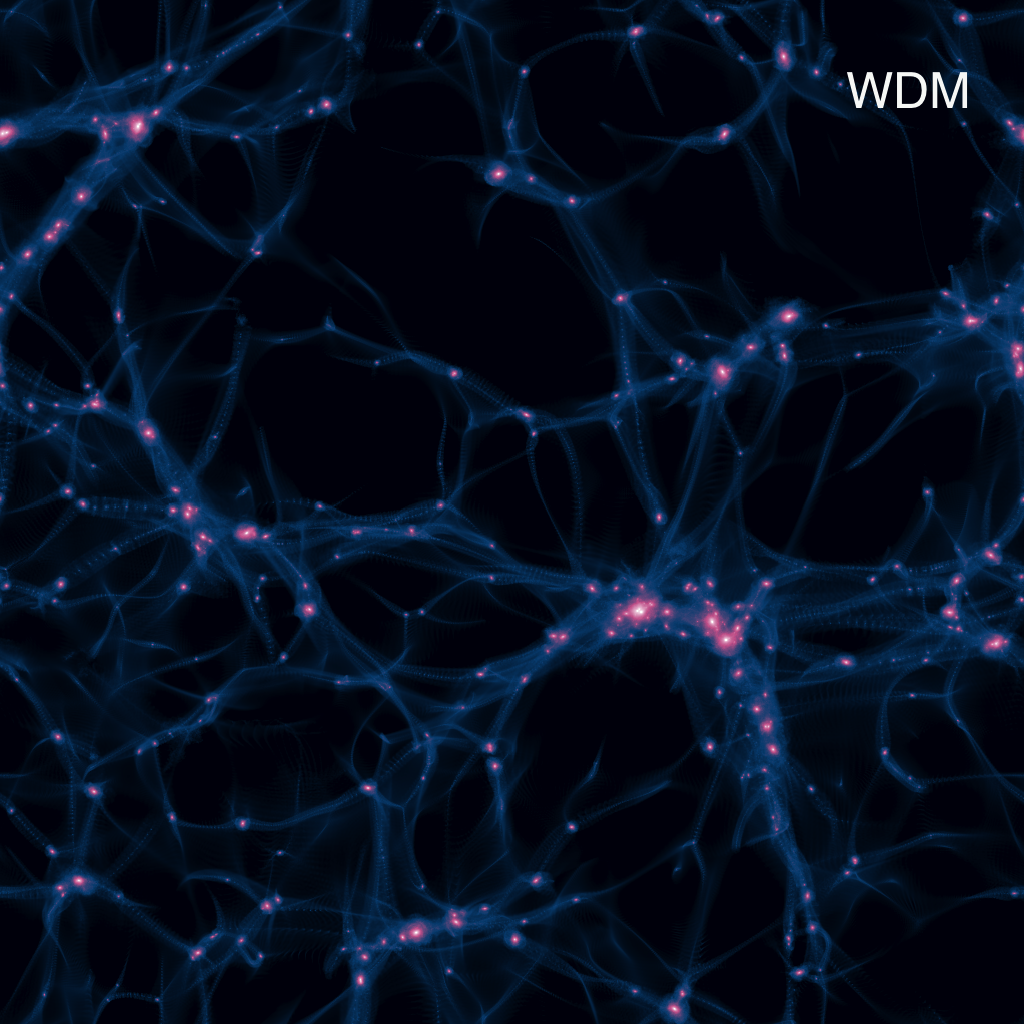}
\caption{Projected dark matter density at $z=0$ in a slice of thickness $20h^{-1}$Mpc through the full box ($64h^{-1}$Mpc on a  side) of four of our simulations, which have $512^3$ particles. Ordered from top left to bottom right, according to the abundance of low-mass halos: CDM, ADM$_\text{sDAO}$, ADM$_\text{wDAO}$ and WDM (see Table \ref{table:simulations}).}
\label{fig:snapshots}
\end{figure}

At larger redshifts, the nonlinear matter power spectra shown in Fig.~\ref{fig:power_spectrum_evol} significantly depart from that of CDM. Indeed, while the power spectra of our three benchmark models were largely indistinguishable at $z=0$ for $k\lesssim 5h$ Mpc$^{-1}$, we note that they all display very different shapes at $z=5$ on the scales probed by our simulations. This indicates that the structure formation history in each model is in general quite different, leading to distinct predictions about the structure of the high-redshift universe as will soon become apparent in our study of the halo mass function. Essentially, low-mass halos ($\lesssim 10^{13}h^{-1}$ M$_{\odot}$) in our two ADM simulations form later than in the CDM case but earlier than in the WDM analogue model, implying that the densities of ADM halos will be somewhat in between these two limiting cases (as, for the moment, we are not discussing the impact of self-scattering). Among other effects, this has implications for the reionization history of the Universe in SIDM models with light mediators and could potentially be probed with high-redshift tracers of the density field such as the 21-cm line \cite{2011MNRAS.413.1174P,2012ApJ...753...81P,Parsons:2013dwa,2013PASA...30...31B,Patil:2014dpa}. 

In Fig. \ref{fig:snapshots}, we give a visual impression of the simulations at $z=0$ by showing the projected dark matter distribution within a slice that is $20h^{-1}$Mpc thick. The color scale is arranged in such a way that regions of higher density appear as bright magenta. The simulations are ordered from top left to bottom right according to their abundance of low-mass halos. We only show four of our simulations since the cases of ADM$_\text{sDAO-env}\text{(nc)}$ and ADM$_\text{sDAO}\text{(nc)}$ are very similar to CDM and ADM$_\text{sDAO}$, respectively. It is already clear at a visual level that the ADM simulations preserve the large-scale structure of CDM but with a deficit of low-mass halos. The case of the WDM simulation is of course more dramatic given the large scale at which the power spectrum has been truncated.

\begin{figure}[t]
\includegraphics[width=0.7\columnwidth]{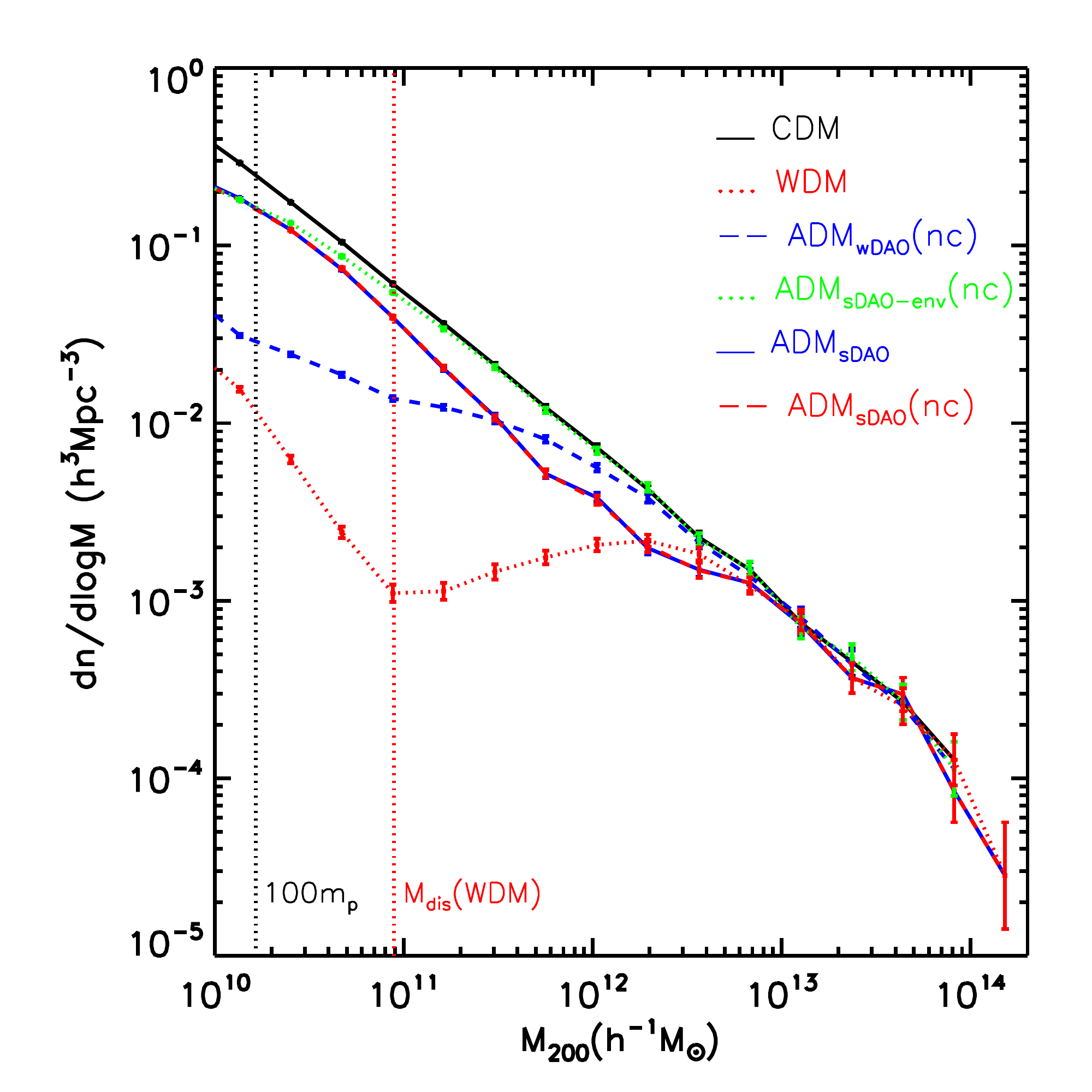}
\caption{ Differential halo mass function (number density of dark matter halos per unit logarithmic mass) as a function of halo mass at $z = 0$ for the simulation suite listed in Table~\ref{table:simulations}. The statistical error bars are Poissonian. 
The clear upturn at $\sim10^{11}h^{-1}$~M$_\odot$ for WDM is due to spurious halos formed due to discreteness effects associated to the sharp
truncation of the power spectrum at small scales. The vertical dotted red line marks this spurious transition for the WDM case using the formula from Ref.~\citep{Wang:2007}. This mass also serves as a {\it conservative} limit for convergence of the mass function for all simulations. The dotted black vertical line marks the mass where halos have 100 particles.}
\label{fig:halospectrum}
\end{figure}

We show in Fig.~\ref{fig:halospectrum} the number density of dark matter halos as a function of halo mass at $z=0$ (differential halo mass function) for dark matter models drawn from the simulation suite described in Table \ref{table:simulations}. We emphasize again that the initial conditions in each simulation (CDM, WDM, ADM$_\text{wDAO}$, {\it etc}.) are the same except for the input linear spectra shown in Fig.~\ref{fig:examplepowerspectrum}. Halos are identified using the friends-of-friends (FOF) algorithm \citep{Davis:1985} with a linking length of $b=0.2$. Afterwards, each FOF halo is searched for self-bound substructures using the \texttt{SUBFIND} algorithm \citep{Springel:2001}. With this algorithm we can identify the center of the gravitational potential for each halo, which we use to construct spherical density profiles. We then define the virial  radius ($r_{200}$) and mass ($M_{200}$) of the halo as the radius where the mean overdensity is 200 times the critical density, and the mass internal to this radius. We caution that these choices imply that, at low masses for the non-CDM models, some of the objects that we define as ``halos'' are in reality structures that are in early stage of collapse or protohalos that are not yet fully virialized \cite{2013MNRAS.434.3337A}. We thus expect the mass functions shown in this section to be conservative upper limits on the actual mass function of virialized dark matter halos. 

Numerical artifacts due to the discreteness of the density field in simulations are prevalent well above the inter-particle separation ($d_{\rm p}$) whenever there is a sharp cut-off in the power spectrum. This situation is well known in WDM $N-$body simulations \citep{Wang:2007,Zavala:2009,Lovell:2014} where spurious halos dominate below a limiting mass of $M_{\rm dis}=10.1\,\Omega_{\rm DM}\rho_{\rm crit}d_{\rm p}k_{\rm peak}^{-2}$, where $k_{\rm peak}$ is the wavenumber where $\Delta^2(k)$ reaches its maximum. We clearly confirm this mass scaling for the WDM case, $M_{\rm dis}\sim10^{11}h^{-1}$~M$_\odot$ (red vertical dotted line in Fig.~\ref{fig:halospectrum}). For our ADM simulations, discreteness effects are an issue at masses at least a factor of a few lower, which is already close to our limit to resolve dark matter halos reliably, as indicated by the vertical black dotted line denoted ``$100m_{\rm p}$'' in Fig.~\ref{fig:halospectrum}. In any case, at all redshifts, we can put a {\it conservative} limit of convergence of the mass function at $M_{200}\sim10^{11}h^{-1}$M$_\odot$ for all simulations (see Appendix).

\begin{figure*}
\begin{centering}
\subfigure{\includegraphics[width=0.33\textwidth]{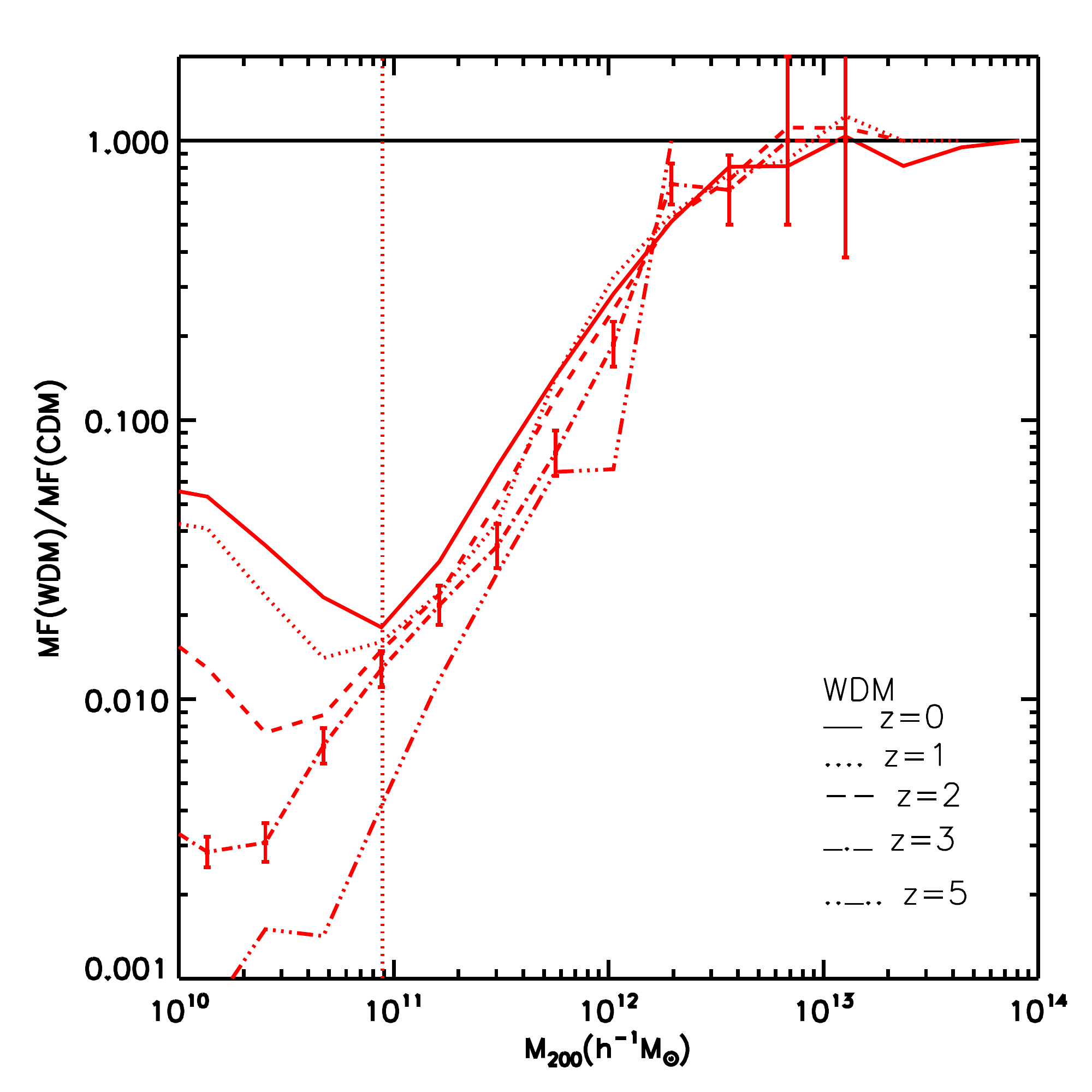}}\subfigure{\includegraphics[width=0.33\textwidth]{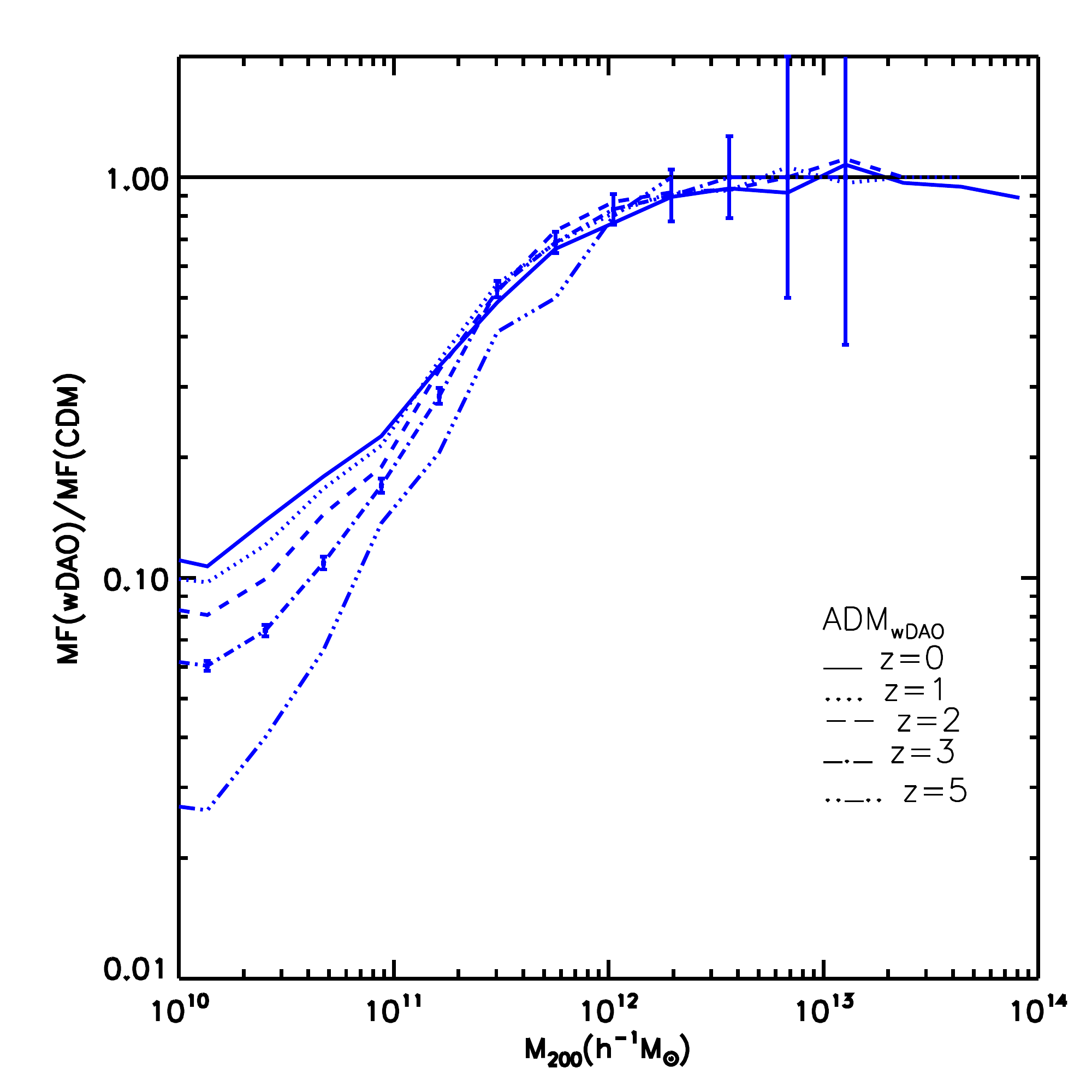}}\subfigure{\includegraphics[width=0.33\textwidth]{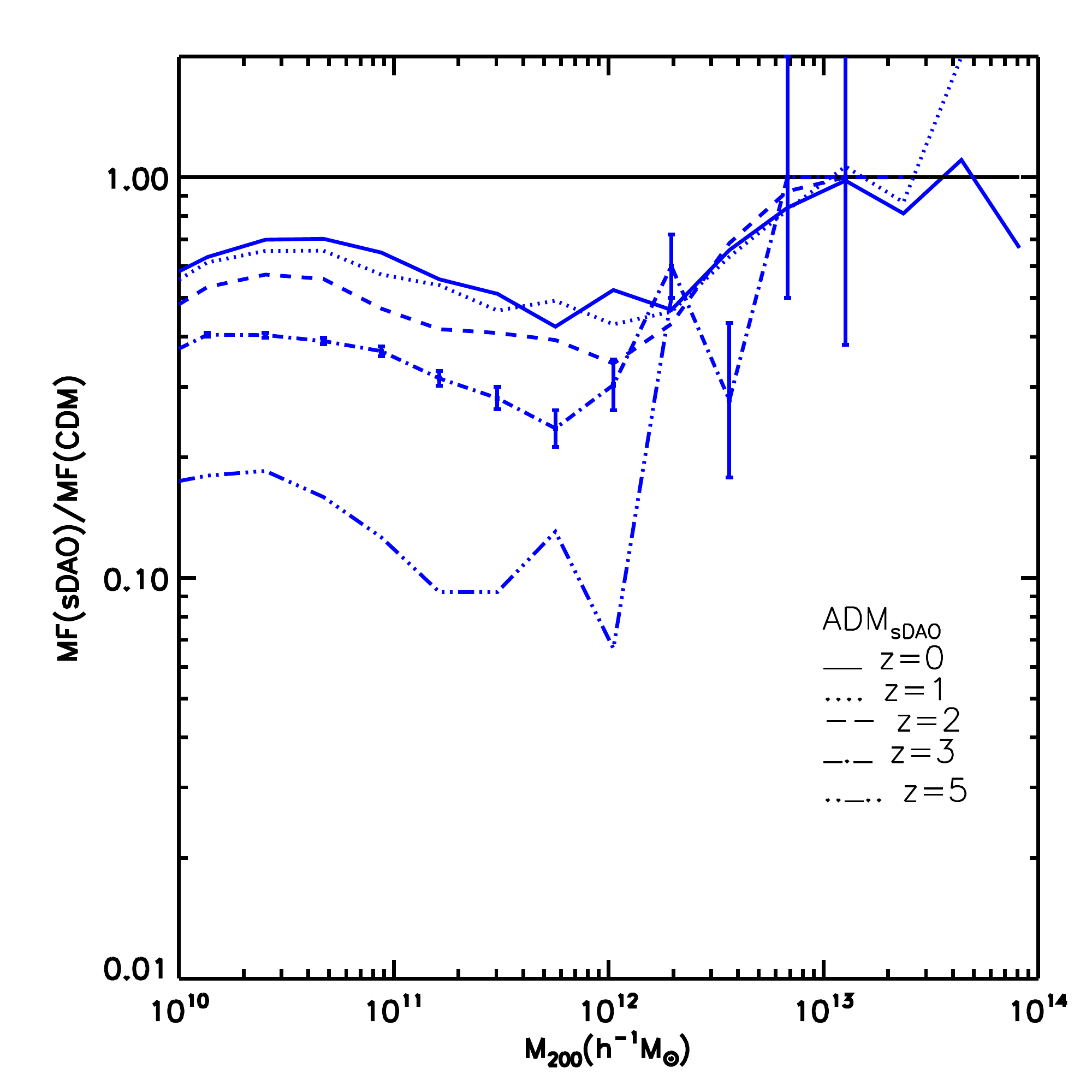}}
\caption{Ratio of the differential mass functions of the WDM(left), ADM$_\text{wDAO}$ (center), and ADM$_\text{sDAO}$ (right) simulations to that of CDM as a function of mass at different redshifts. Statistical error bars are shown for $z=3$ for reference purposes. The vertical dotted line in the left panel marks the mass where spurious halos due to numerical artifacts dominate the mass function (same as in Fig. \ref{fig:halospectrum}). This mass, $\sim10^{11}h^{-1}$M$_\odot$, is also an effective convergence mass for all our simulations.}
\label{fig:mf_evolution}
\end{centering}
\end{figure*}

The CDM simulation serves as the default scenario against which the WDM and ADM models can be compared. We show in Fig.~\ref{fig:mf_evolution} the ratio of the halo mass functions of the WDM (left panel), ADM$_\text{wDAO}$ (middle panel), and ADM$_\text{sDAO}$ (right panel) simulations to that of CDM over a range of redshifts. Several things are immediately apparent from Figs.~\ref{fig:halospectrum} and \ref{fig:mf_evolution}. The WDM simulation has the largest deviation from the CDM baseline, starting at high halo masses ($\sim 4\times10^{12}h^{-1}$M$_\odot $ at $z=0$). The suppression of the halo masses in the strong DAO scenario appears at similarly high halo masses, but critically, more halos in this mass range are formed in this simulation than in either the WDM or weak DAO cases. The weak DAO simulation has large suppression of halos formed at masses smaller than $\sim 10^{12}h^{-1}$M$_\odot$. 

This pattern is as expected given the initial power spectra shown in Fig.~\ref{fig:examplepowerspectrum}. WDM has essentially no power on small scales in the initial matter power spectrum, which results in large suppression of low-mass halos in the early Universe. This is borne out in the simulations: looking at the evolution of the WDM mass function over redshift (left panel, Fig.~\ref{fig:mf_evolution}), we see that low mass halos are not initially present in large numbers, but develop later. The low-mass halos form later than would occur in CDM; as the average dark matter density in the Universe is lower at the time of formation, the resulting low-mass halos have lower central densities than one would expect in CDM. A similar effect occurs in the ADM models, The impact on halo density profiles will be explored in more detail in Section~\ref{sec:density}.

Our model of ADM with strong DAO has a power spectrum which deviates from CDM at the same wavenumber as our WDM benchmark. This is reflected in the $z=0$ mass spectrum as both simulations have suppressed halo abundance below the same mass scale. However, due to the acoustic oscillations, the power is not exponentially suppressed at larger wavenumbers, and so the initial suppression of low-mass halos is not as severe as with WDM, although their number density never fully reaches that of the CDM benchmark. The formation of small structures is delayed relative to CDM, although not as dramatically as in the WDM case. This implies that ADM$_{\rm sDAO}$ halos with masses corresponding to scales between $r_{\rm DAO}$ and $r_{\rm SD}$ are typically denser than their WDM counterparts and thus less likely to be tidally disrupted in the later Universe (although self-scattering changes this picture somewhat as we show in the next section). 

Moreover, we observe in Fig.~\ref{fig:mf_evolution} that the overall shape of the ADM$_{\rm sDAO}$ halo mass function departs significantly from that of either WDM or ADM$_{\rm wDAO}$ whose deviation from CDM is monotonically increasing towards low halo masses. We instead see that the initial suppression of the mass function around the scale corresponding to $r_{\rm DAO}$ ($M_{200}\sim 4\times10^{12}h^{-1}$ M$_\odot$) is followed by an increase at masses below 
$M_{200}\lesssim 3\times10^{11}h^{-1}$ M$_\odot$, before becoming suppressed again on scales affected by Silk damping. This behavior of the ADM$_{\rm sDAO}$ mass function, which is apparent over the broad range of redshifts displayed in Fig.~\ref{fig:mf_evolution}, seems to indicate that the two distinct scales characterizing SIDM models with a light force carrier can remain imprinted in the mass distribution of objects populating the Universe. While further work is required, it is interesting that such non-trivial structures set by the early cosmology of the SIDM models could survive to the present day.

The strong DAO envelope simulation (ADM$_\text{sDAO-env}$ in Fig.~\ref{fig:halospectrum}) shows no reduction of the mass function until near $10^{11} h^{-1}$M$_\odot$, which is in line with our expectation as its initial matter power spectrum is not exponentially suppressed until $k \sim 10h$~Mpc$^{-1}$. This scale is also where we expect the acoustic oscillations of the strong DAO model to be severely damped. We indeed observe that the mass functions of our ADM$_\text{sDAO-env}$ and ADM$_{\rm sDAO}$ simulations become similar for masses below  $\sim2\times10^{10} h^{-1}$M$_\odot$, in line with our expectations. Unfortunately, halos at this mass scale are poorly resolved in our numerical experiments and further work is needed to unambiguously determine that the suppression at these scales is indeed caused by Silk damping of the initial dark matter density field. We also note that the collisional and non-collisional simulations of ADM$_{\rm sDAO}$ have identical late-time nonlinear matter power spectra. This is expected, as the self-scattering cross section of this model is too small to significantly reduce the number of massive halos. While non-zero $\sigma_\text{tr}$ can evaporate small halos, those are far below the resolution limit of our simulations. 

The weak DAO simulation shows deviations from the CDM predictions which become significant on slightly smaller mass scales compared to ADM$_\text{sDAO}$ ($\sim 10^{12}h^{-1} $M$_\odot$). Again, this is as expected from the initial matter power spectrum, which begins to diverge from CDM at wavenumbers of a few $h/$Mpc. This deviation starts small and increases significantly at lower ($\sim 3 \times 10^{11}h^{-1}M_\odot$) halo masses, although not as quickly as in the WDM case. Since our weak DAO model is characterized by $r_\text{SD} \sim r_\text{DAO}$, we would naturally expect the acoustic oscillations to play a subdominant role (compared to the strong DAO case) in preventing a strong suppression of the mass function. Looking at Fig.~\ref{fig:mf_evolution} and comparing the WDM and ADM$_\text{wDAO}$ cases, we observe a milder suppression of the mass function in the weak DAO case, indicating that the severely damped acoustic oscillations in this model indeed have an important effect. 

In summary, we see that the rich phenomenology of SIDM coupled to a relativistic force carrier in the early Universe leads to observables that are both qualitatively and quantitatively different than both WDM and CDM. The nonlinear evolution of the matter power spectrum in each model leads to a unique structure formation history and overall behavior of the halo mass function. For the first time, we have characterized the shape of the mass function of SIDM models coupled to a light mediator and shown it to have features that cannot be easily mimicked by a warm or cold dark matter scenario. 

\section{Inner halo densities: effect of Self-Interactions \label{sec:density}}

We now turn from the abundance of dark matter halos to their internal structure. In this paper, we examine only the dark matter density profiles, leaving additional properties such as velocity distributions for future work. As mentioned previously, SIDM models with long range forces can modify the dark matter profiles in two possible ways. The first is the (by now well known) effect of non-zero $\sigma_\text{tr}/m_D$, which allows for energy transport within the dark matter halo itself. Energy transfer from collisions in the high density central region can transform a cuspy NFW-like profile \cite{Navarro:1995iw} into a cored profile. As only the ADM$_\text{sDAO}$ simulation was performed with collisional dark matter, this is the only set of halos in which this effect could be seen. 
\begin{figure*}[t!]
\begin{centering}
\subfigure{\includegraphics[width=0.45\textwidth]{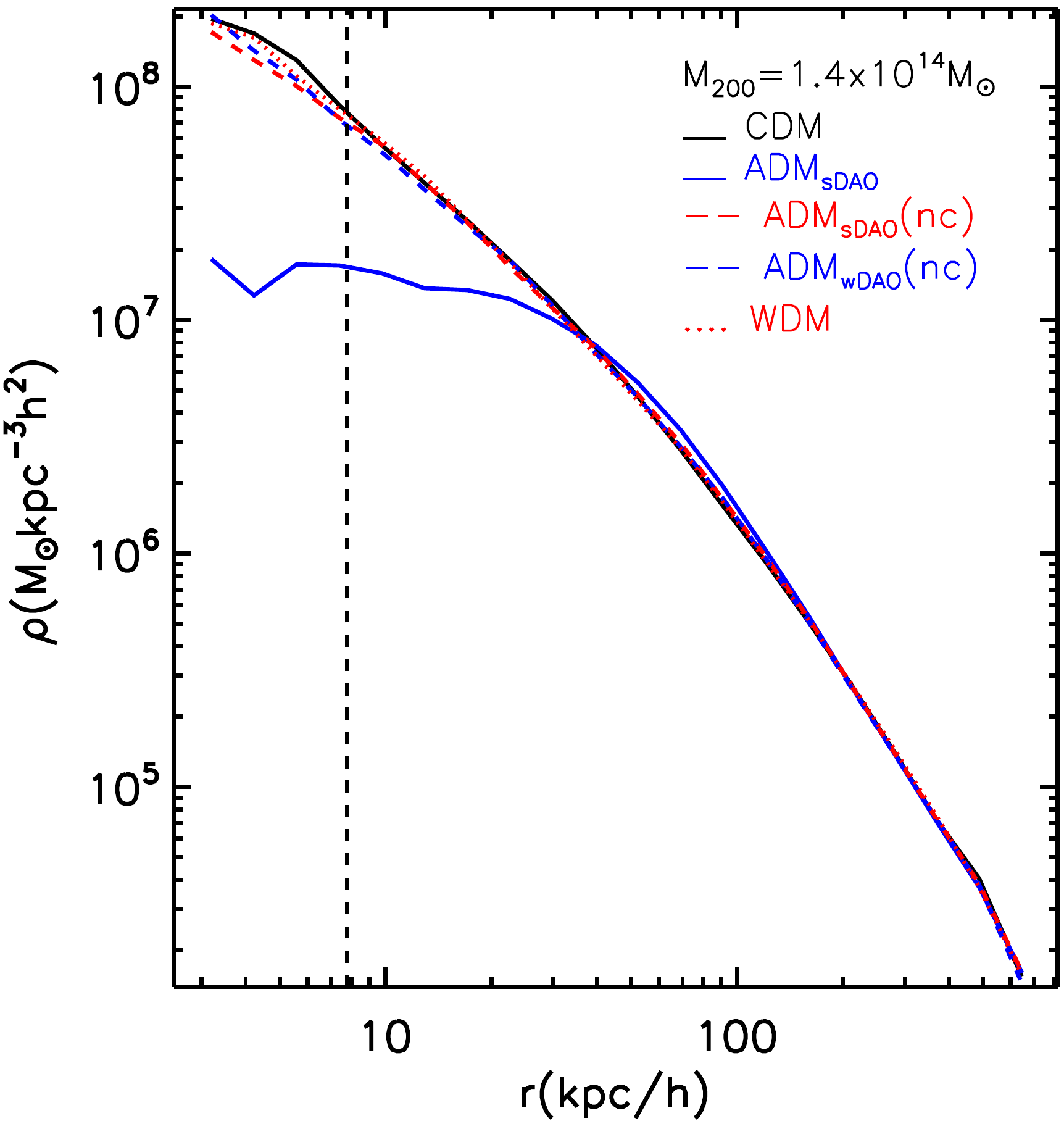}}\subfigure{\includegraphics[width=0.45\textwidth]{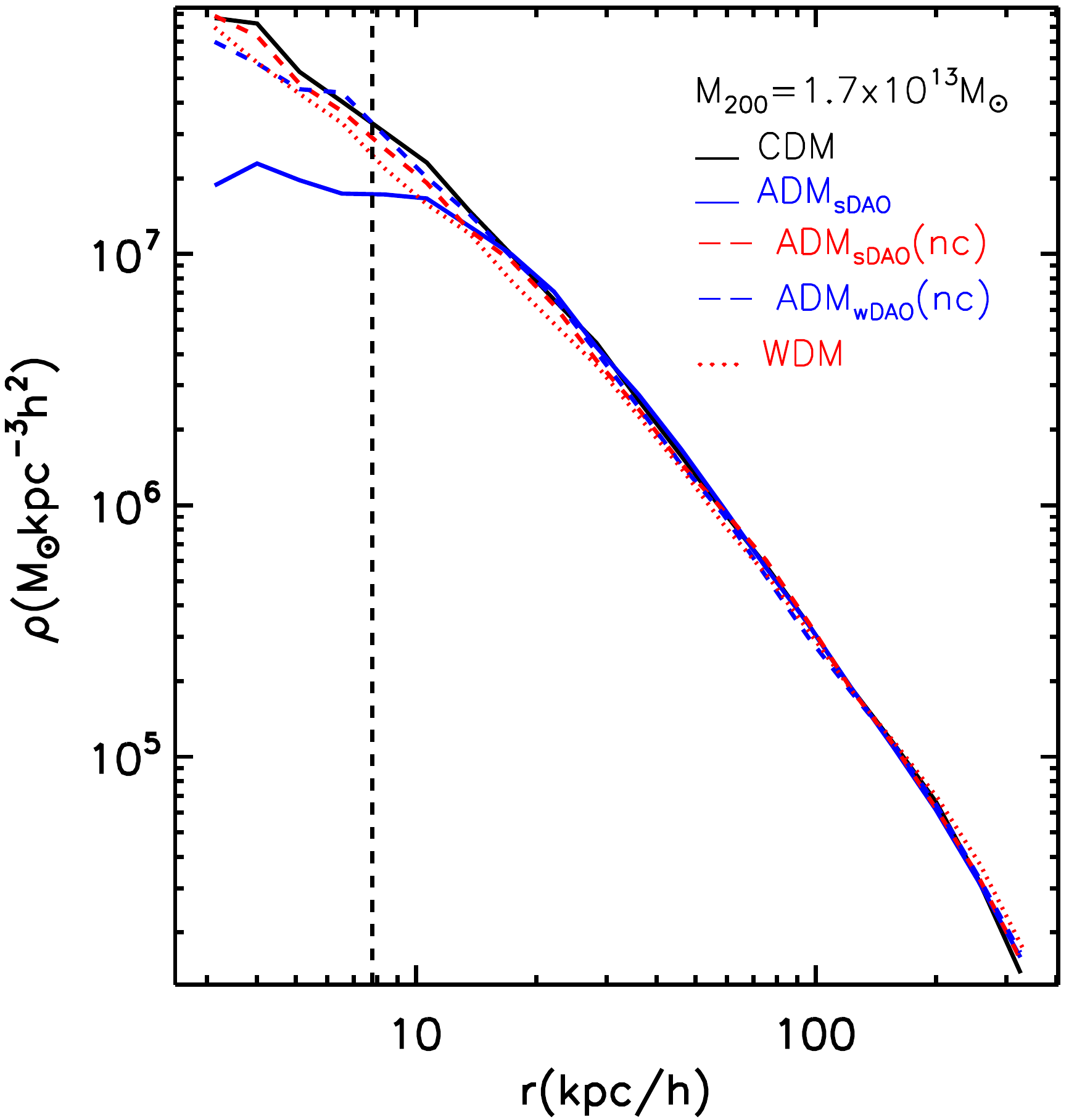}}

\subfigure{\includegraphics[width=0.45\textwidth]{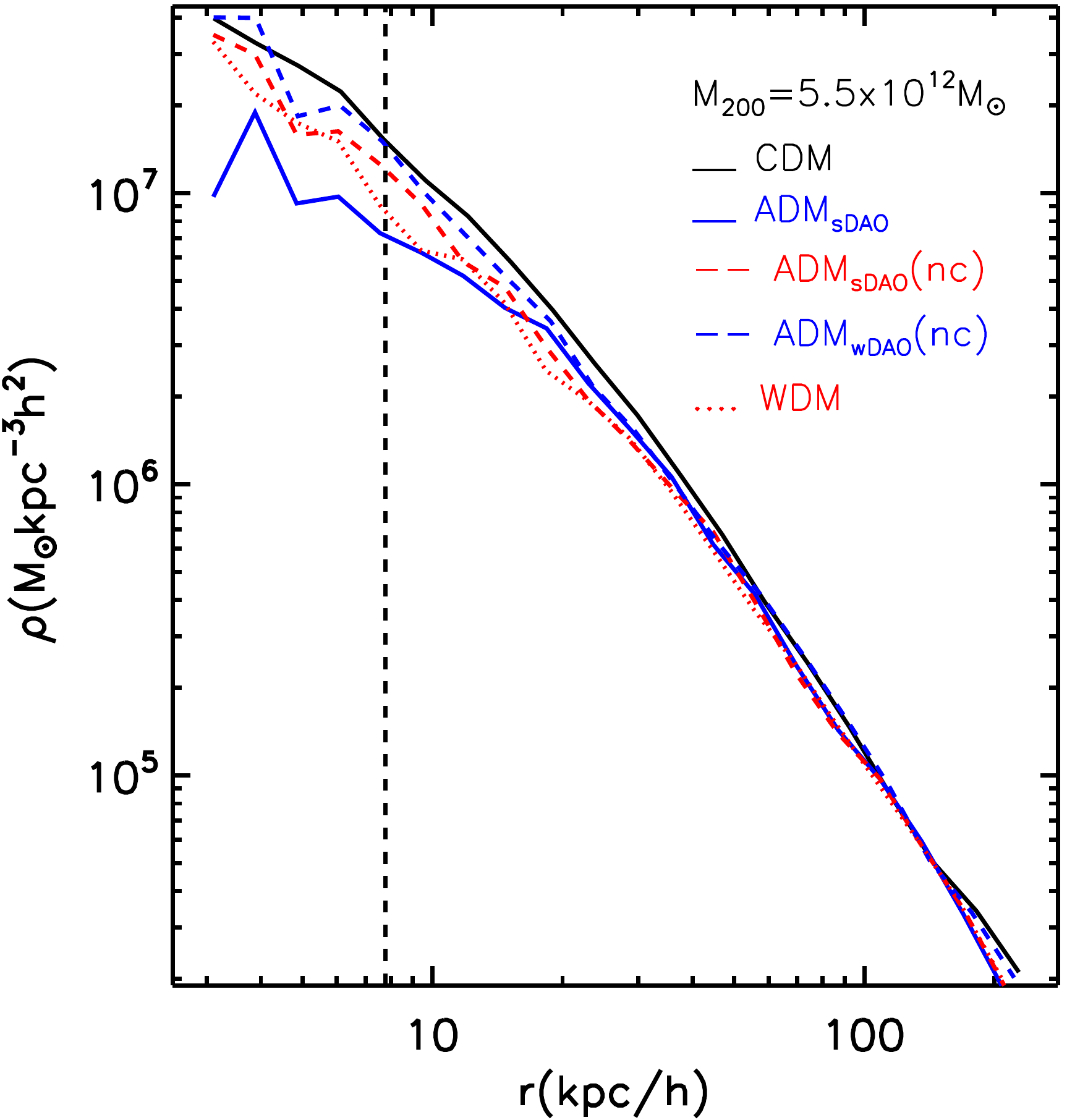}}\subfigure{\includegraphics[width=0.45\textwidth]{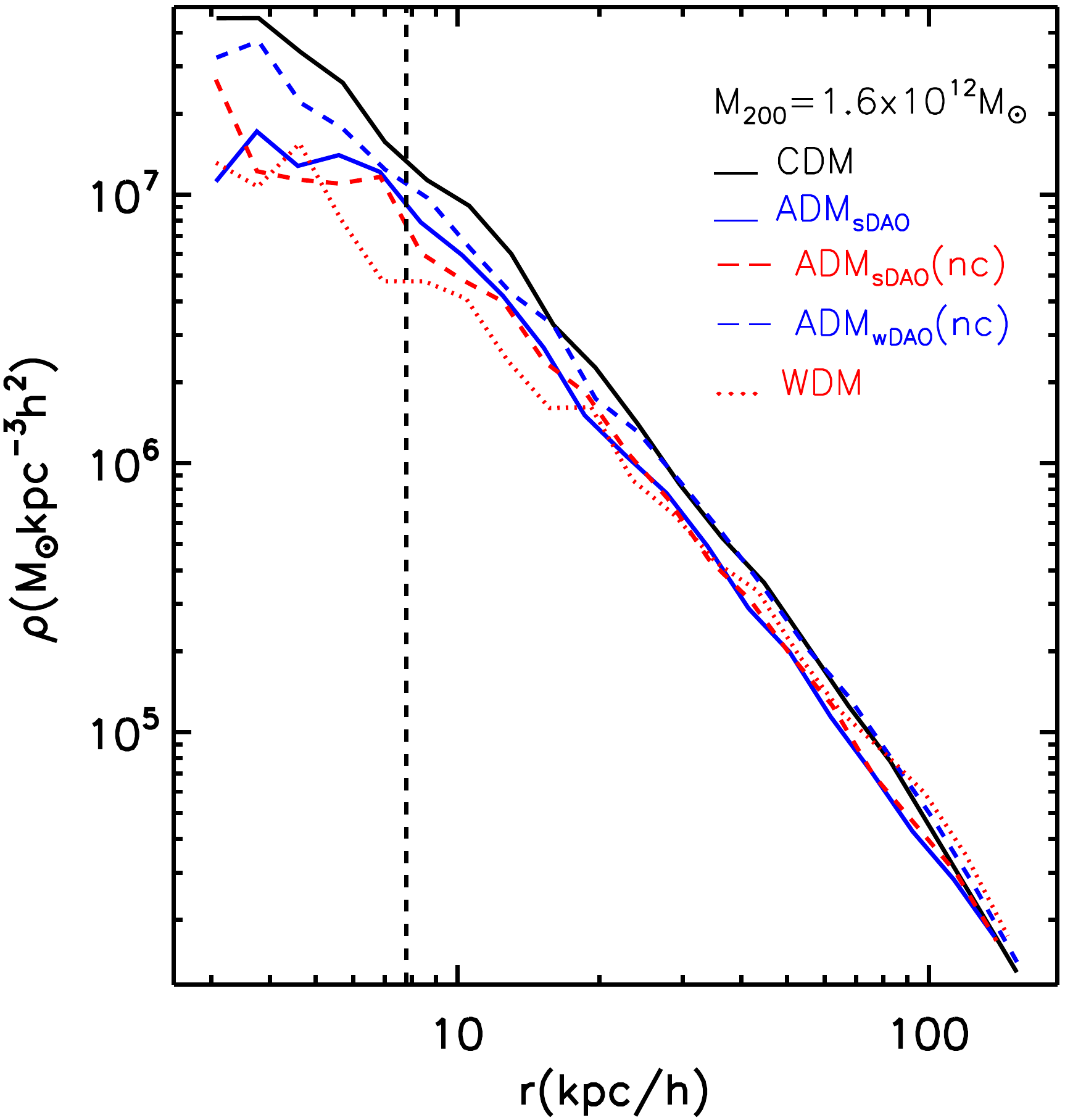}}
\caption{Radial density profiles for four example halos at $z=0$ with masses of $M = 1.4\times 10^{14}$, $1.7 \times 10^{13}$, $5.5\times 10^{12}$, and $1.6\times 10^{12}M_\odot$ (from top left to bottom right). The vertical dashed line marks the resolution limit $3\epsilon$ of our simulation, where $\epsilon \sim 2.8h^{-1}$~kpc is the Plummer-equivalent softening length. Notice how only the simulation with self-interactions develops cored dark matter profiles, clearly resolved at the highest masses. The remaining simulations have less dense low-mass (close to the filtering mass scale) halos than CDM  due to the modification of their initial power spectrum. Although less dense than CDM, these halos still have a steep NFW-like inner density profile. The value of the virial mass in each legend is that of the CDM simulation.}
\label{fig:dens_profiles}
\end{centering}
\end{figure*}

In addition, the initial suppression of small scales in the WDM and ADM models results in low-mass halos forming later in the Universe's evolution, as mentioned in the previous section. As the dark matter density is lower at later times, this delay in halo formation at small scales results in less dense halos. However, this does not change the shape of the density profiles, resulting in the low-mass halos being less dense, but still NFW-like \citep[e.g. for WDM,][]{Avila-Reese:2001}. 

The combination of the two ways of modifying the dark matter density profiles can be seen in Fig.~\ref{fig:dens_profiles}, where we plot the dark matter density as a function of radius for halos of mass $1.4 \times 10^{14}$, $1.7 \times 10^{13}$, $5.5 \times 10^{12}$ and $1.6 \times 10^{12}M_\odot$. As can be seen, the collisional simulation ADM$_\text{sDAO}$ has a profile that deviates significantly from a cuspy NFW, {\it specially} at the largest halo masses. As the non-collisional ADM$_\text{sDAO}$ simulation shows no such modification for the highest mass halos, this effect can be unambiguously associated with the dark matter self-interaction. Though this self-interaction modification of the profile appears to extend down to the smallest mass halos, the radius at which the deviation from NFW occurs drops below the the resolution of our simulations.

Looking now at the lower mass halos shown in Fig.~\ref{fig:dens_profiles}, we see that those formed by the simulations with a large exponential suppression in the primordial power spectrum (WDM, ADM$_\text{wDAO}$, and ADM$_\text{sDAO}$ in both collisional and non-collisional runs) have suppressed densities when compared to CDM. This effect is absent for the $1.4\times 10^{14}M_\odot$ halos. Note that this highest mass range corresponds to wavenumbers of $k < 1~h/$Mpc, where all the simulations have power spectra that are identical to CDM. These halos are therefore assembled at the same time as in the CDM benchmark, which is reflected in their density profiles. The WDM simulation, with the greatest suppression of power in the initial conditions, is seen to have the greatest reduction in dark matter density in the present-day low-mass halos. Both the collisional and non-collisional ADM$_\text{sDAO}$ runs have a similar reduction of density at this mass scale, although only the collisional run has the deviation from NFW profiles in the inner slope of halos, as discussed previously.

\begin{figure}
\includegraphics[width=0.6\textwidth]{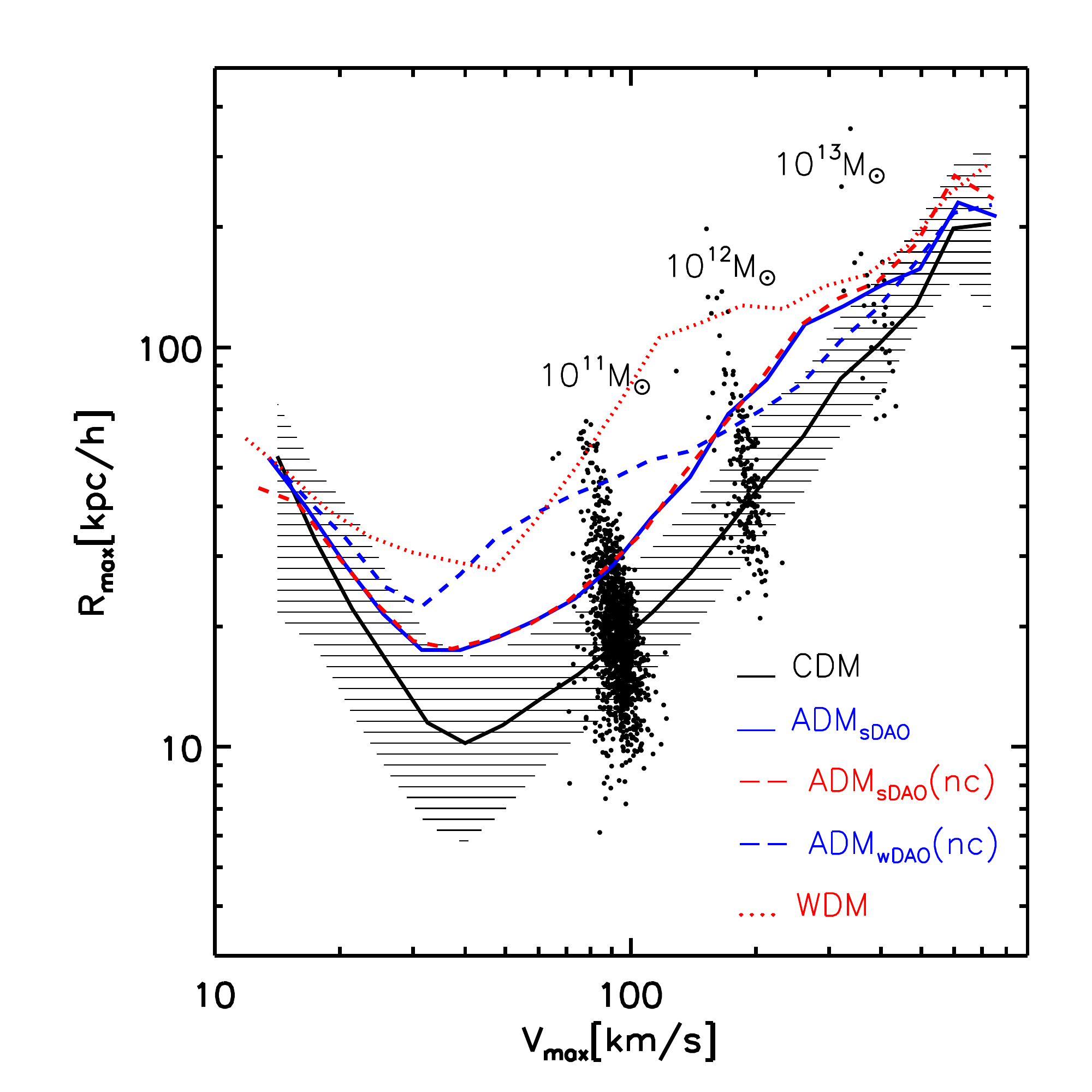}
\caption{Plot of the maximum circular velocity ($V_{\rm max}$) versus the radius at this maximum ($R_{\rm max}$) for all {\it main} (central) halos for the different simulations. For the CDM case, we show the median and $1\sigma$ regions of the distribution (shaded area). All CDM halos with masses around $10^{11}$M$_\odot$, $10^{12}$M$_\odot$ and $10^{13}$M$_\odot$ are shown with small dots. The other lines are the medians of the distributions in each simulations (the spread is similar to CDM in all cases). The upturn at lower velocities,  $V_{\rm max}\sim50$~km/s, is created by resolution effects.}
\label{fig:inner_structure}
\end{figure}

This reduction in halo density in simulations with suppressed small scale power (relative to CDM) can be seen statistically in Fig.~\ref{fig:inner_structure}. Here, for each halo within a simulation, we plot the maximum velocity $V_\text{max}$ (a measure of the total enclosed mass) versus the radius $R_\text{max}$ at which this maximum velocity is found. The black solid line and shader area show the median and $1\sigma$ regions of the CDM distributions. For the other simulations, only the median is drawn. This figure demonstrates more clearly that the concentration of halos from simulations with suppressed primordial power spectra is much lower than that predicted by CDM. An effect caused by the delayed formation of smaller halos. Again, we see the largest deviation for the WDM scenario, where the power spectrum is suppressed the most at small scales. 

\begin{figure}
\includegraphics[width=0.6\textwidth]{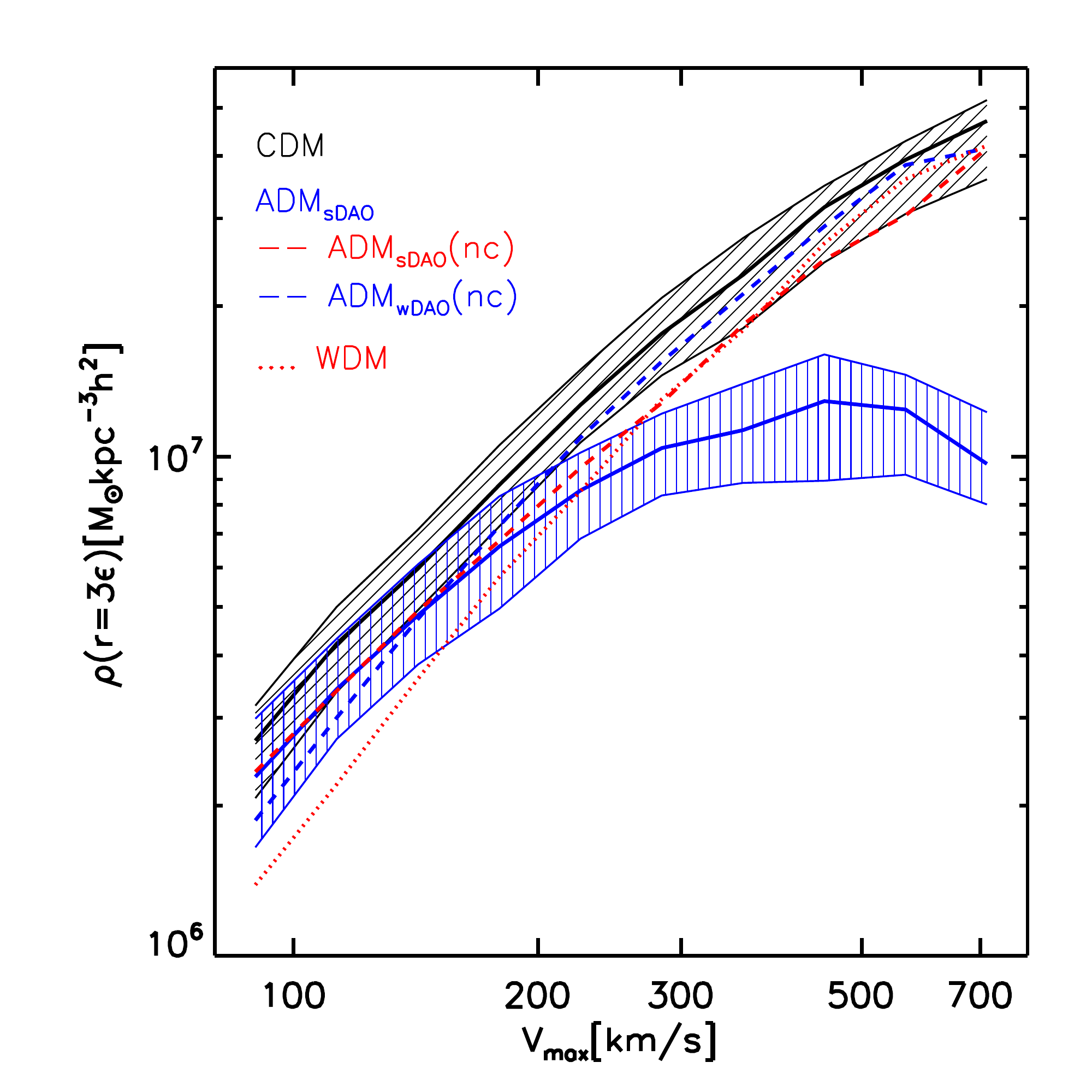}
\caption{Dark matter density at a halo-centric radius
of $3\epsilon$, where $\epsilon$ is the Plummer-equivalent softening length of the simulation, as a function of the maximum rotational velocity $V_{\rm max}$ of the halo. This radius is roughly the radius down to which we can trust our simulations. The thick lines are the median of the distributions for each case. For the CDM and ADM$_{\rm sDAO}$ cases we also show the $1\sigma$ region of the distributions; the spread for the other cases is similar. A given value of $V_{\rm max}$ roughly correspond to a given virial mass (see Fig. \ref{fig:inner_structure}). }
\label{fig:inner_density}
\end{figure}

More importantly, notice that in Fig.~\ref{fig:inner_structure} the effect of self-interactions is not visible (compare the collisional and non-collisional ADM$_\text{sDAO}$ lines). This is because the bulk of the collisions occur well within $R_{\rm max}$. The effects of self-interactions are apparent only at inner radii as shown in Fig.~\ref{fig:inner_density}, where we plot the density at a fixed radius versus $V_\text{max}$. The central radius chosen is $R = 3\epsilon \sim 8~h^{-1}$kpc, where the density profiles for all cases studied here are sufficiently converged (see Appendix). Although all non-CDM simulations are less dense than CDM at this radius, the effect of collisions is the most dramatic for most massive halos. Again, the models with the greatest initial suppression of power on small scales have the greatest reduction in density. Furthermore, the addition of self-interactions in the ADM$_\text{sDAO}$ simulation significantly reduces the central densities relative to the non-collisional run. 

 We note that given the scale of the power spectrum cut-off of the WDM case and assuming a thermal relic scenario, the impact of the primordial thermal velocities of the dark matter particles might be relevant to structure formation and evolution. We have not however attempted to include such velocities since the proper implementation scheme is still unclear. This is not expected to be important for the halo mass function since the typical collapse velocities are much larger than the thermal velocity dispersion in the case we have considered. Our WDM model is actually close to the one presented in Ref.~\cite{2013MNRAS.434.3337A}, where this statement is discussed quantitatively. The inner WDM halo densities might still be affected by thermal velocities (developing a dark matter core, see {\it e.g.}~Refs.~\cite{Avila-Reese:2001,Maccio:2012}). However, the region where this is important is expected to be below the resolution of our simulations, {\it i.e.}, for the WDM case we have studied here, the expected core is much smaller than the one developed in the ADM$_\text{sDAO}$ case (see Ref.~\cite{Maccio:2012}). Moreover, as we mentioned before, it is still controversial how to properly include thermal velocities in simulations.

\section{Conclusions}

Though not necessary for a model of self-interacting dark matter (SIDM), the large transfer cross sections needed to significantly alter the profile of dark matter halos can easily be realized via a new light-mediator force coupling to the dark matter. If this force carrier is light enough ($\lesssim$ MeV), then it acts as a long-range force during the early Universe, keeping the dark matter and dark radiation in kinetic equilibrium with a relativistic sound speed. This introduces a collisional damping scale into the dark sector ($r_{\rm SD}$), which would suppress the formation of halos below a critical size, set by the dark sector parameters. Furthermore, the dark radiation bath allows for dark acoustic oscillations which introduces an additional scale, the sound horizon of dark matter ($r_{\rm DAO}$), into the initial matter power spectrum. As the DAO scale depends on a different combination of dark sector parameters, the relationship between the two key scales ($r_{\rm DAO}$ and $r_{\rm SD}$) can vary greatly across models. In addition, the non-zero momentum transfer cross section between dark matter particles in the present day can significantly alter the density profiles of dark matter halos and, if large enough, reduce the number density of low-mass halos.

Though previous works \cite{Mangano:2006mp,Hooper:2007tu,Kaplan:2009de,Shoemaker:2013tda} have commented on these effects individually, no full cosmological simulations have been performed until now. In this paper, we performed an initial exploration of the phenomenology available to SIDM models with a long range force, using the dark atom model as a benchmark (ADM). Though we restricted ourselves to relatively large dark matter halo masses, we have demonstrated that the multiple scales inherent in this class of models can lead to observable modifications of the halo properties, which are distinct from either cold dark matter (CDM) or warm dark matter (WDM) scenarios. In particular, we have demonstrated that the imprint of the Silk (collisional) damping scale and the DAO scale can survive in the differential halo mass spectrum to $z=0$. Furthermore, even without the effects of non-zero scattering cross section, the density profiles of low mass halos are altered, as the delayed collapse results in a suppression of the inner halo densities. With sufficiently large self-interactions, the inner slopes of the density profiles can also be modified; turning NFW-like halos into cored-like systems.

The suppression of halos on smaller mass scales could be probed by future strong lensing studies \cite{Dalal:2001fq,Fadely:2011su,Keeton:2008gq,Koopmans:2009mq,Marshall:2009eu,Moustakas:2008ib,Moustakas:2009na,Vegetti:2008eg,Vegetti:2009cz,Vegetti:2009gw,Vegetti:2012mc}. As can be seen, SIDM with long range force carriers can modify the mass spectrum away from CDM in a quantitatively different manner than WDM. As seen in Fig.~\ref{fig:mf_evolution}, echoes of the initial DAO structure are visible in the mass functions of a strong DAO ADM model 
even at low redshifts. 
With future lensing measurements probing the low-mass halo regime, it may be possible to not only discriminate WDM models from CDM, but also to find evidence for long-range dark mediators through their effects on the multi-epoch halo mass function.

In this first exploratory paper, we have restricted ourselves to studying SIDM scenarios which alter relatively large ($\sim 10^{11-12}M_\odot$) halos. As such, we do not address in detail the potential of such models to resolve the various outstanding crises in small-scale structure faced by CDM \cite{Klypin:1999uc,Moore:1999nt,Spergel:1999mh,Zavala:2009,Oh:2010ea,BoylanKolchin:2011dk,BoylanKolchin:2011de,Walker:2011zu,2013MNRAS.433.3539G,Weinberg:2013aya,Kirby:2014sya,Tollerud:2014zha,Garrison-Kimmel:2014vqa}. Straightforward extrapolation of our results to smaller values of $r_\text{SD}$ and $r_\text{DAO}$ would indicate that the addition of long-range forces to models of self-interaction would, in addition to the well-known formation of dark matter cores in low-mass haloes \cite{Vogelsberger:2012ku,Zavala:2012us,Rocha:2012jg}, also reduce the abundance of low-mass halos, while leaving larger objects relatively unchanged. However, further work is needed to confirm this intuition, likely requiring higher resolution simulations, as well as the important addition of velocity-dependent cross sections and baryonic physics. Given the novel effects of these long-range force models, such models and simulations are likely to provide unique and interesting cosmologies that can be directly compared and constrained by observations in the near future.

\begin{center}
{\bf Acknowledgments} 

MRB would like to thank Alyson Brooks for useful discussion and comments. MRB and JZ thank the Aspen Center for Physics, where the initial conversations that led to their interest in this project were held. The Aspen Center for Physics is supported by the National Science Foundation under Grant 1066293.  
The work of F.-Y. C.-R. was performed in part at the California Institute of Technology for the Keck Institute for Space Studies, which is funded by the W. M. Keck Foundation. Part of the research described in this paper was carried out at the Jet Propulsion Laboratory, California Institute of Technology, under a contract with the National Aeronautics and Space Administration.  The research of KS is supported
in part by a National Science and Engineering Research
Council (NSERC) of Canada Discovery Grant.
The Dark Cosmology Centre is funded by the DNRF. JZ is supported by the EU under a Marie Curie International Incoming Fellowship, contract PIIF-GA-2013-627723. 

\end{center}

\appendix*

\section{Convergence Tests}

To show that the results we have presented are not strongly affected by the resolution of our simulations, we performed convergence tests for the halo mass function and for the halo density profiles. For each of the simulations we ran a complimentary set of identical simulations with a factor of 8 fewer particles than used in our primary results, {\it i.e.}, $256^3$ particles rather than $512^3$. With such a test we can establish {\it conservative} levels of convergence for our higher resolution simulations.
\begin{figure*}[h!]
\begin{centering}
\subfigure{\includegraphics[width=0.45\textwidth]{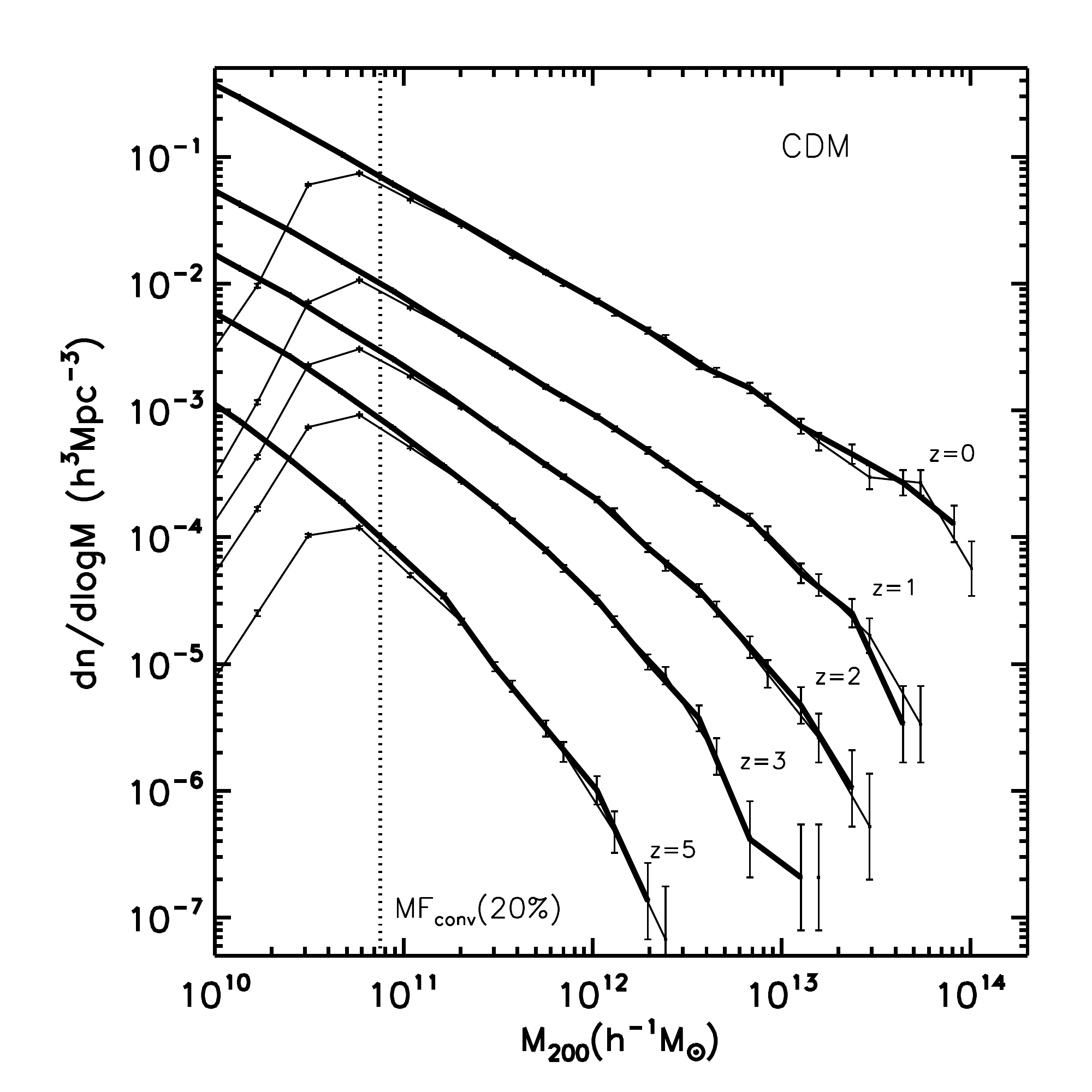}}\subfigure{\includegraphics[width=0.45\textwidth]{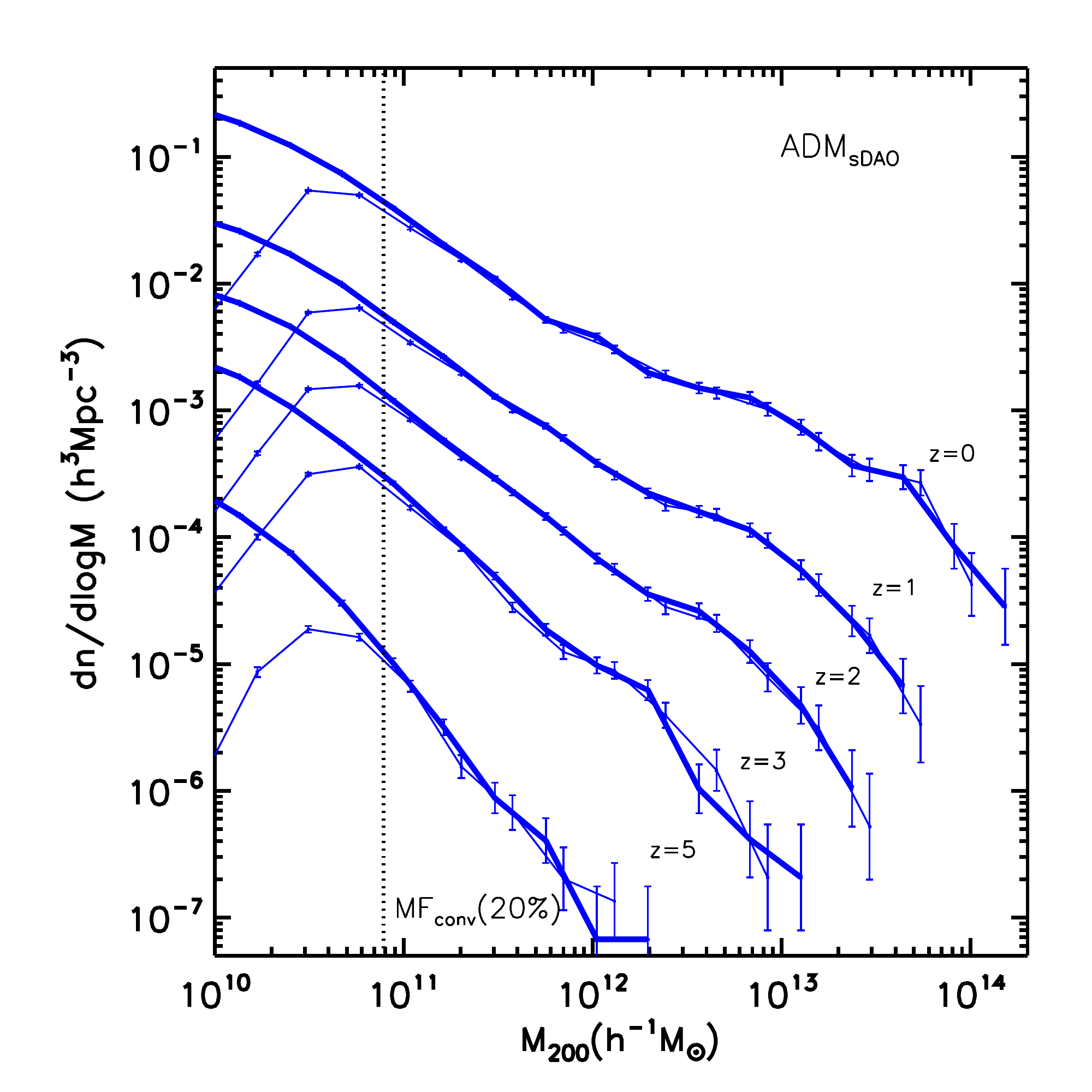}}

\subfigure{\includegraphics[width=0.45\textwidth]{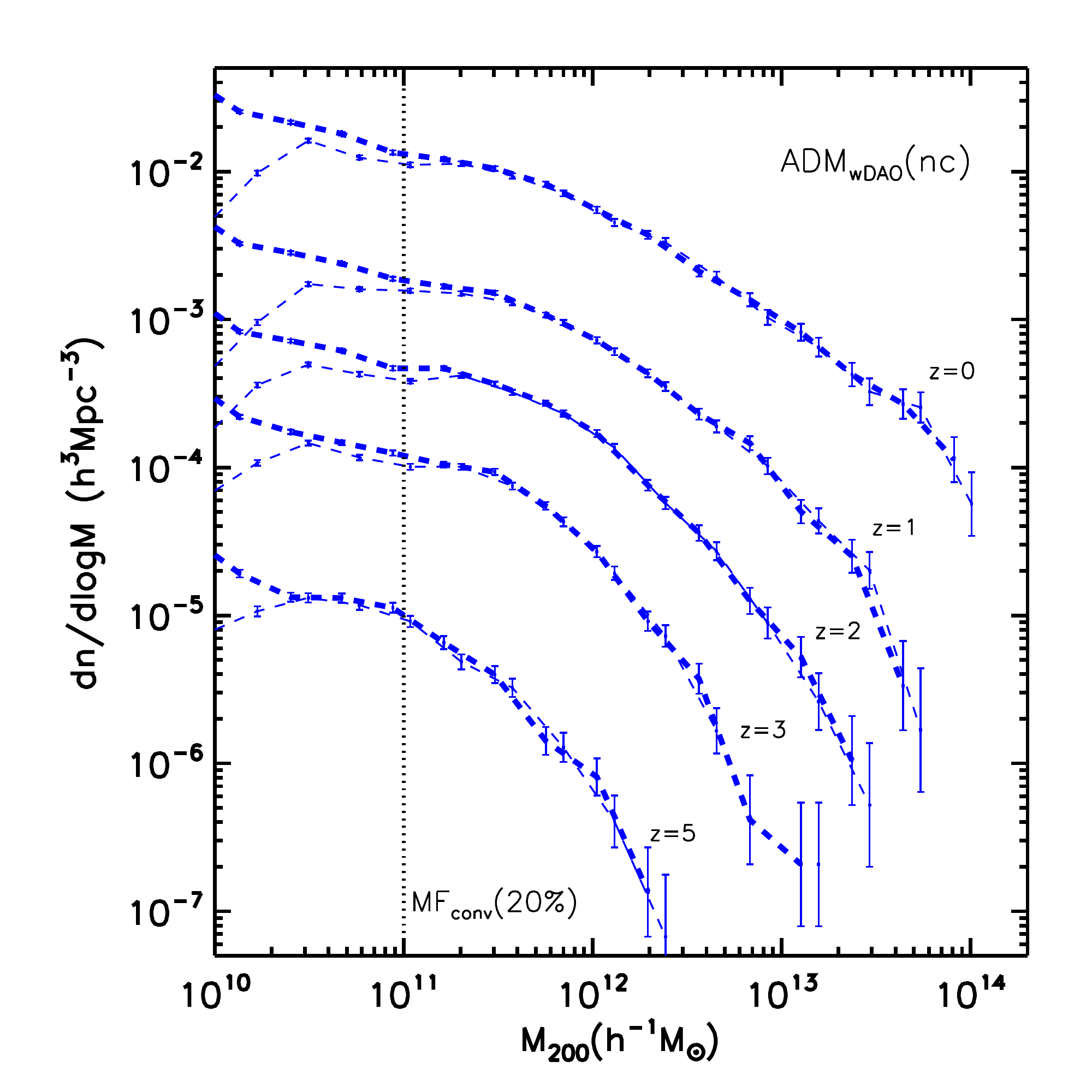}}\subfigure{\includegraphics[width=0.45\textwidth]{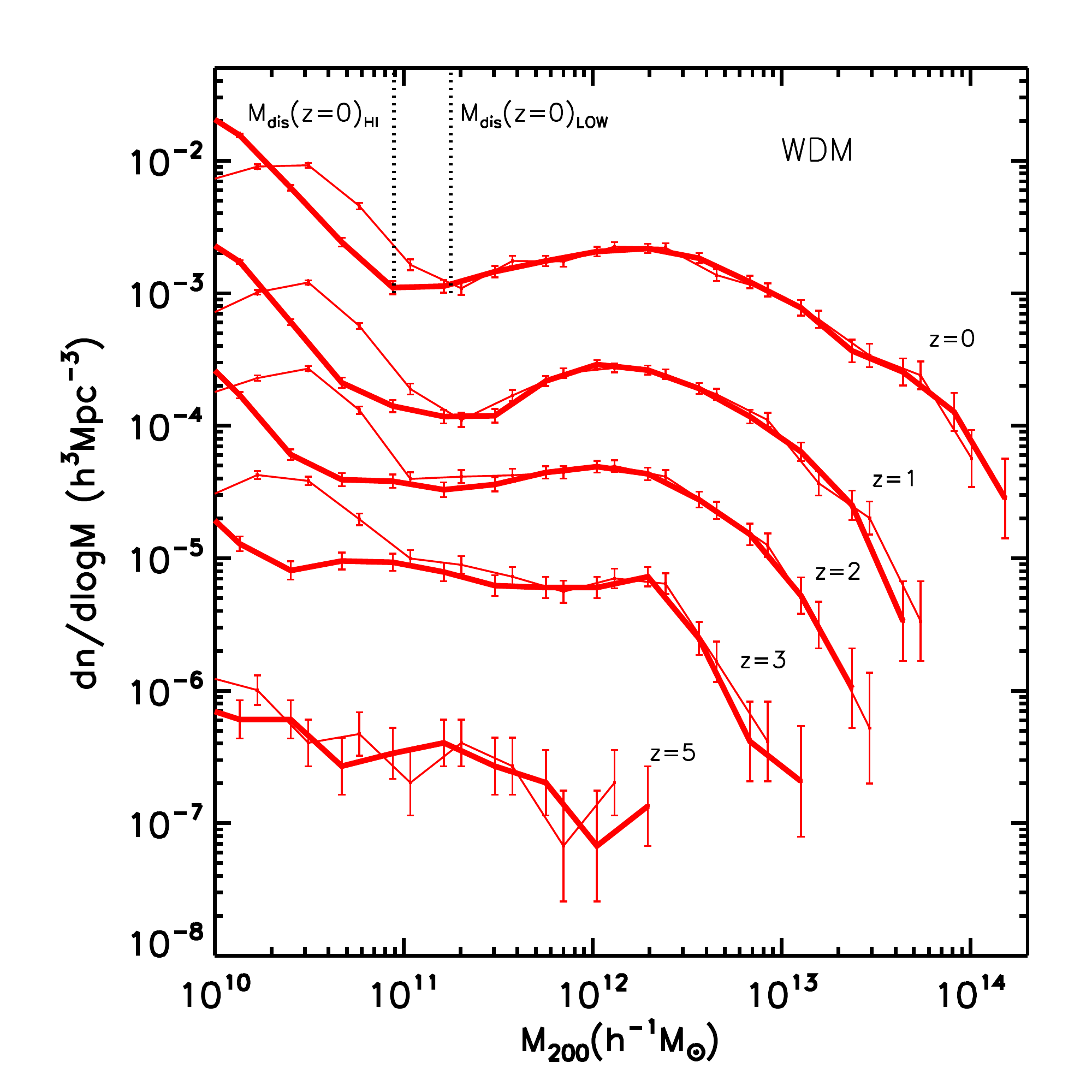}}
\caption{Differential halo mass function (number density of dark matter halos per unit logarithmic mass) as a function of halo mass at different redshifts for four of our simulations according to the legend (see Table~\ref{table:simulations}). Our highest
resolution levels ($512^3$ particles) are shown with thick lines while lower resolution versions ($256^3$ particles) of the same simulations are shown with thiner lines. The Poissonian statistical error bars are shown for both resolution levels. For the simulations: CDM (top left), ADM$_\text{sDAO}$ (top right), and ADM$_\text{wDAO}$ (bottom left), we mark the value of $M_{200}$ where the convergence of the
mass function, {\it at the lowest resolution level}, is better than $20\%$ at all redshifts.  In the case of the WDM simulation (bottom right), the vertical dotted lines mark the masses where discreteness effects are important (at $z=0$) according to the formula given by Ref.~\citep{Wang:2007}. Our simulations are clearly consistent with the upturn expected at these masses due to spurious halos. Overall, all our high resolution simulations have a mass function that is converged at $M_{\rm 200}\sim10^{11}~h^{-1}$M$_\odot$, which is sufficient for the main conclusions of our paper.}
\label{fig:MF_convergence}
\end{centering}
\end{figure*}

Fig.~\ref{fig:MF_convergence} shows the differential halo mass function at different redshifts for the main simulations in our work: all simulations which appear in Table~\ref{table:simulations} with the exceptions of ADM$_\text{sDAO}\text{(nc)}$, which has a mass function almost identical to  ADM$_\text{sDAO}$, and ADM$_{\rm sDAO-env}$(nc), which is very similar to CDM, except at the smallest masses ($M_{200}\lesssim10^{11}~h^{-1}$M$_\odot$).
The high (low) resolution level is shown with a thick (thin) line. Except for the WDM case, even the low resolution simulations are essentially converged at $M_{200}\sim10^{11}~h^{-1}$M$_\odot$. This is quantified with the vertical dotted lines, which show the value of the virial mass where the convergence in the mass function is $\lesssim20\%$ at all redshifts. At larger masses, errors in the mass function are essentially dominated by counting statistics. The vertical lines are therefore quite conservative limits of the virial mass ($M_{200}$) where the mass function is not affected by numerical resolution for our highest resolution runs. In the case of the WDM simulations, convergence is harder to achieve due to the strong impact of discreteness effects. In these cases the minimum mass we can trust is reliably given by the formula given in Ref. \cite{Wang:2007}, which implies a mass resolution improvement that scales as $N^{1/3}$. For our high resolution simulation, this limiting mass is $\lesssim10^{11}~h^{-1}$, while for the low resolution version is a factor of 2 larger. These expectations appear with dotted lines in the upper left of the WDM panel in Fig.~\ref{fig:MF_convergence}. Our simulations clearly confirm that this is the appropriate limit of convergence for the mass function.

We therefore conclude that the results presented in this paper regarding the mass function are numerically converged at $M_{\rm 200}\sim10^{11}~h^{-1}$M$_\odot$ at all redshifts. Except for the WDM simulation, this is a {\it conservative} limit of convergence, that is nevertheless sufficient to support our main conclusions. 

\begin{figure*}[t!]
\begin{centering}
\subfigure{\includegraphics[width=0.44\textwidth]{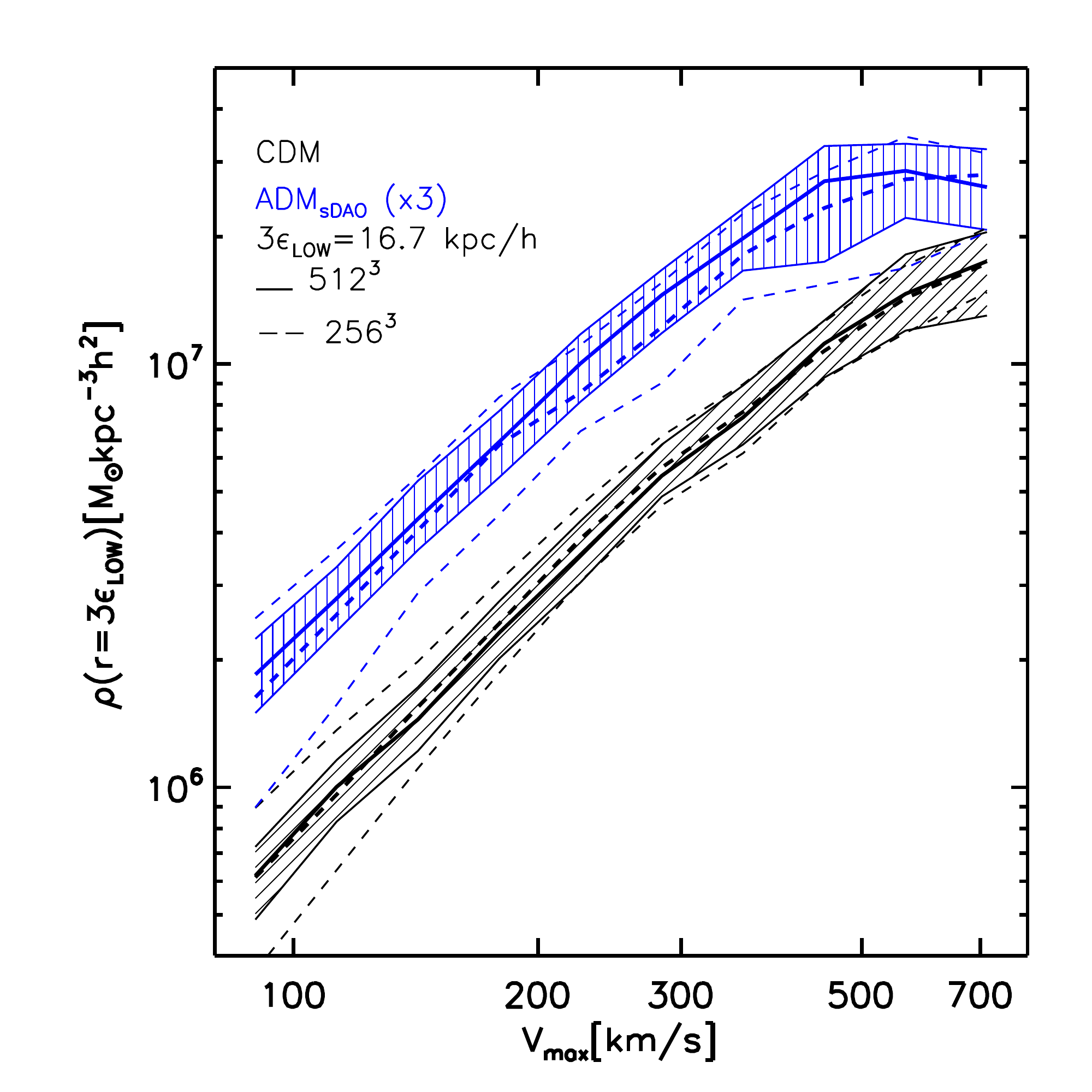}}\subfigure{\includegraphics[width=0.44\textwidth]{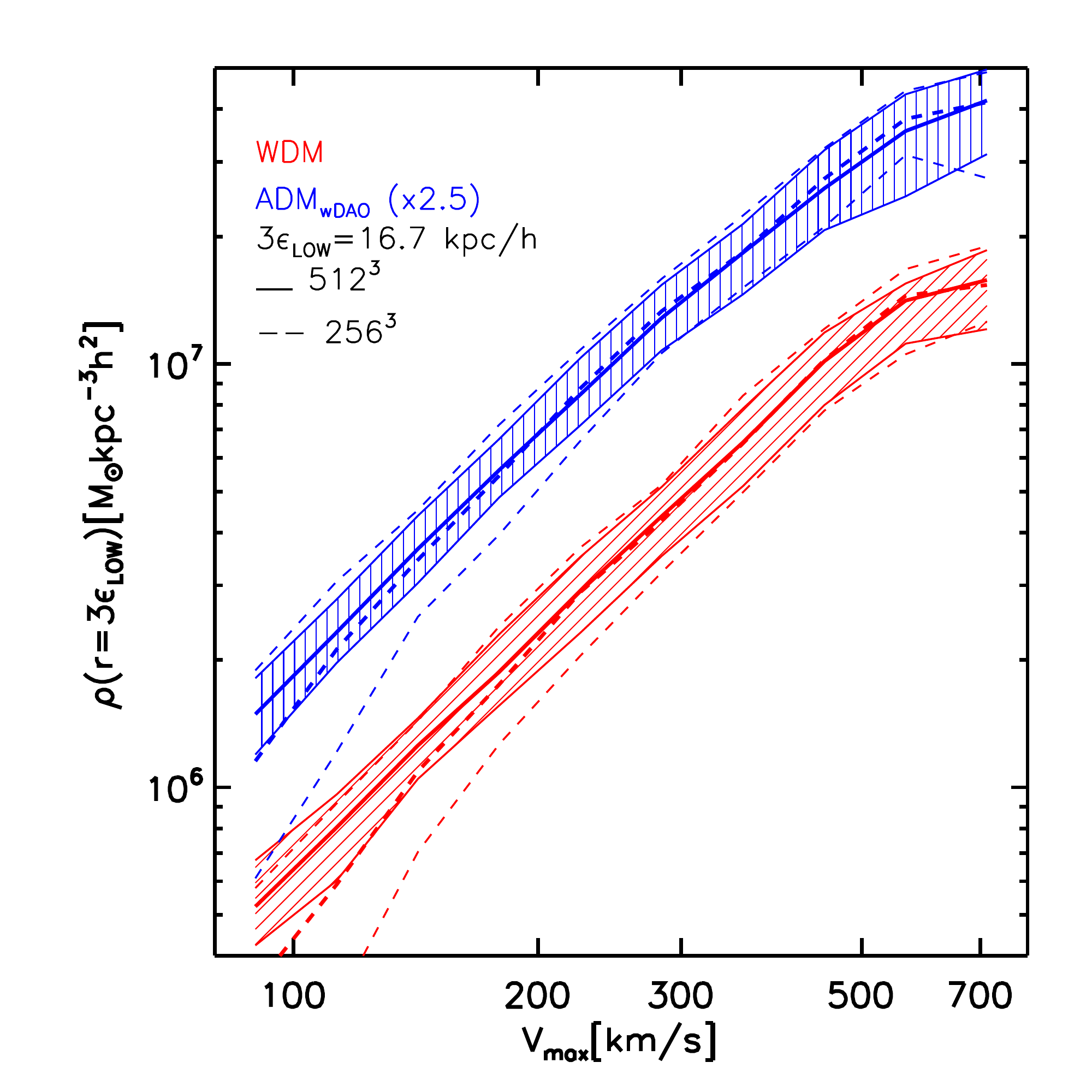}}
\caption{Dark matter density at a fixed radius ($3\epsilon_{\rm LOW}$) from the center of each halo as a function of its maximum rotational velocity $V_{\rm max}$. The Plummer-equivalent softening length for our low resolution simulations ($256^3$ particles), $\epsilon_{\rm LOW}$, is $5.6~h^{-1}$~kpc. We show the results for four of our simulations, two in each panel: CDM and ADM$_\text{sDAO}$ in the left and WDM and ADM$_\text{wDAO}$ in the right. For clarity, the cases of ADM$_\text{sDAO}$ and ADM$_\text{wDAO}$ have been shifted vertically by a factor of 3, 2.5, respectively. The high (low) resolution case is shown with solid (dashed) lines. We show the median and $1\sigma$ region of the distributions for each case (shown also as shaded areas for the high resolution case). The convergence in the median of the distributions is within the $1\sigma$ statistical spread for all cases, except at lowest $V_{\rm max}$ (mass) values for WDM. }
\label{fig:dens_convergence}
\end{centering}
\end{figure*}

Fig.~\ref{fig:dens_convergence} shows a statistical convergence test for the inner density of dark matter halos at a fixed radius of three times the Plummer-equivalent softening length of the low resolution simulation: $3\epsilon_{\rm LOW}=16.7h^{-1}$kpc. This radius was chosen so that we can assess if the densities extracted from the high resolution simulations are to be trusted at the corresponding radius of $3\epsilon_{\rm HI}=8.4h^{-1}$kpc, which is the radius we use in Fig.~\ref{fig:inner_density}. The plot shows the distribution (median and $1\sigma$ regions) of the densities at a radius $3\epsilon_{\rm LOW}$ as a function of the maximum circular velocity $V_{\rm max}$ of the halos in the high (solid lines) and low (dashed lines) resolution simulations. Four dark matter models are shown: CDM and ADM$_\text{sDAO}$ (shifted vertically up by a factor of 3) in the left and WDM and ADM$_\text{wDAO}$ (shifted vertically up by a factor of 2.5) in the right. We can see that for CDM, the densities are essentially converged at this radius across all the mass ($V_{\rm max}$) range, although at the low end, the spread in the distribution is clearly larger in the low resolution case. This is because the values of $V_{\rm max}$ and $R_{\rm max}$ start to be affected importantly by resolution at $V_{\rm max}\sim200$~km/s for the low resolution simulations. The WDM and ADM$_\text{wDAO}$ show a similar level of convergence as CDM in this plot, except at the low end where, for a fixed value of $V_{\rm max}$, the central densities are systematically underestimated in the low resolution case. In the case of ADM$_\text{sDAO}$, there is a small underestimation of the central densities across all scales in the low resolution case, which signals that the inner densities have not fully converged at $3\epsilon_{\rm LOW}$. However this underestimation is minimal, clearly less than the $1\sigma$ statistical spread  of the distribution. The lack of full convergence in the ADM$_\text{sDAO}$ case is caused by the fact that a poorly resolved initial density cusp of a halo has an impact on the collision frequency that is used in the algorithm for self-scattering. SIDM simulations (with a CDM power spectrum), have also shown that inner halo densities are only minimally affected at a radius of $3\epsilon$ for simulations where the spatial resolution varies by a factor of $\sim6$ \citep[see Fig. 9 of][]{Vogelsberger:2012ku}. 

We are therefore confident that for all our simulations, the dark matter densities at a radius of $3\epsilon$ are reliable, particularly for high mass halos, which is the regime where we draw our main conclusions.

\bibliographystyle{apsrev}
\bibliography{sidm}

\end{document}